%% file: main.tex
\documentclass[sigconf,letterpaper,10pt]{acmart}

\usepackage{markdown}
\usepackage{xspace}
\usepackage{float}
\usepackage{graphicx}
\usepackage[linesnumbered,noline,noend,figure]{algorithm2e}
\usepackage[aboveskip=0pt,belowskip=0pt,labelfont={bf,small},textfont={it,small}]{subcaption}
\usepackage[aboveskip=0pt,belowskip=0pt,labelfont={bf,small},textfont={it,small}]{caption}
\usepackage{tikz}
\usetikzlibrary{fit,calc}
\usepackage{xcolor}
\usepackage{amsmath}
\usepackage{etoolbox}
\usepackage{setspace}
\usepackage{hyperref}
\usepackage[yyyymmdd,hhmmss]{datetime}
\usepackage{times}

\input{macros}

\acmDOI{}
\acmISBN{}
\acmYear{}
\copyrightyear{}
\acmPrice{}
\renewcommand\footnotetextcopyrightpermission[1]{}
\setcopyright{none}
\begin{document}

\title[Localizing Router Configuration Errors Using MCSes]{Localizing Router Configuration Errors Using Minimal Correction Sets}


\author{{\rm Aaron Gember-Jacobson, Ruchit Shrestha, Xiaolin Sun}\\
Colgate University\\
{\tt \{agemberjacobson,rshrestha,xsun\}@colgate.edu}}

\begin{abstract}
Router configuration errors are unfortunately common and difficult to localize using current network verifiers. We introduce a novel configuration error localizer (CEL) that precisely identifies which configuration segments contribute to the violation of forwarding requirements. In particular, CEL generates a system of satisfiability modulo theories (SMT) constraints---which encode a network's configurations, control logic, and forwarding requirements---and uses a domain-specific minimal correction set (MCS) enumeration algorithm to identify problematic configuration segments. CEL efficiently locates several configuration errors in real university networks and identifies all routing-related and at least half of all ACL-related errors we introduce.

\end{abstract}

\maketitle

\input{intro}
\input{techniques.tex}
\input{approach.tex}
\input{encoding.tex}
\input{mcses.tex}
\input{evaluation.tex}

\input{related.tex}
\input{conclusion.tex}

\pagebreak

\bibliographystyle{abbrv}
\bibliography{main}

\appendix
\input{encoding_appendix}

\end{document}

%% file: macros.tex
\newcommand{\minisection}[1]{\smallskip\noindent{\bf #1.}}
\newcommand{\minisectionpreamble}[1]{\smallskip\noindent{\bf #1}}

\newcommand{\secref}[1]{\S\ref{#1}}
\newcommand{\secsref}[2]{{\S\ref{#1} and \S\ref{#2}}}

\newcommand{\figref}[1]{Figure~\ref{#1}}
\newcommand{\figsref}[2]{{Figures~\ref{#1} and \ref{#2}}}
\newcommand{\tabref}[1]{Table~\ref{#1}}
\newcommand{\appref}[1]{Appendix~\ref{#1}}

\AtBeginEnvironment{algorithm}{\small}
\SetEndCharOfAlgoLine{}
\SetArgSty{}
\SetKwComment{Comment}{//\ }{}
\SetCommentSty{itshape}

\newcommand{\incoming}[3]{\ensuremath{\text{#1}_{{#2} \leftarrow {#3}}}}
\newcommand{\outgoing}[3]{\ensuremath{\text{#1}_{{#2} \rightarrow {#3}}}}
\newcommand{\valid}{\ensuremath{\text{valid}}}
\newcommand{\prefix}{\ensuremath{\text{prefix}}}
\newcommand{\cost}{\ensuremath{\text{cost}}}
\newcommand{\best}[2]{\ensuremath{\text{#1}_{{#2}-\text{best}}}}

\newcommand{\destination}{\ensuremath{\text{destination}}}
\newcommand{\source}{\ensuremath{\text{source}}}

\newcommand{\true}{\ensuremath{\tt True}}
\newcommand{\false}{\ensuremath{\tt False}}

\newcommand{\config}[2]{\ensuremath{\text{cfg}^\text{#1}_{#2}}}
\newcommand{\configOspf}[2]{\config{OSPF:{#1}}{#2}}
\newcommand{\configOspfFilterDefault}[1]{\configOspf{filter}{#1:\text{default}}}
\newcommand{\configOspfOriginate}[2]{\configOspf{originate}{#1:#2}}
\newcommand{\configOspfCost}[2]{\configOspf{cost}{{#1} \rightarrow {#2}}}
\newcommand{\configOspfAdjacency}[2]{\configOspf{adjacency}{{#1} \rightarrow {#2}}}
\newcommand{\configAcl}[2]{\config{ACL}{{#1}:{#2}}}
\newcommand{\configOutAcl}[2]{\config{OutACL}{{#1} \rightarrow {#2}}}

\newcommand{\failedLink}[2]{\ensuremath{\text{failed}_{{#1} \leftarrow {#2}}}}

\newcommand{\mcs}{\ensuremath{\gamma}}
\newcommand{\mss}{\ensuremath{\kappa}}
\newcommand{\mus}{\ensuremath{\mu}}
\newcommand{\genericSubset}{\ensuremath{\sigma}}
\newcommand{\allConstraints}{\ensuremath{\varphi}}
\newcommand{\genericConstraint}[1]{\ensuremath{x_{#1}}}
\newcommand{\configConstraints}{\ensuremath{C}}
\newcommand{\configConstraint}[1]{\ensuremath{c_{#1}}}
\newcommand{\logicConstraints}{\ensuremath{L}}
\newcommand{\logicConstraint}[1]{\ensuremath{l_{#1}}}
\newcommand{\requireConstraints}{\ensuremath{R}}
\newcommand{\requireConstraint}[1]{\ensuremath{r_{#1}}}
\newcommand{\failedConstraints}{\ensuremath{F}}
\newcommand{\failedConstraint}[1]{\ensuremath{f_{#1}}}

\newcommand{\surveycount}{25\xspace}

\RequirePackage{color, soul}

\ifdefined\commentenabled
  \newcommand{\aaron}[1]{{\color{white}\sethlcolor{magenta}\textbf{\hl{[AGJ: #1]}}}}
  \newcommand{\ruchit}[1]{{\color{red}\textit{[RS: #1]}}}
  \newcommand{\owen}[1]{{\color{blue}\textit{[OS: #1]}}}
\else
  \newcommand{\aaron}[1]{}
  \newcommand{\ruchit}[1]{}
  \newcommand{\owen}[1]{}
\fi

\newcommand{\name}{CEL\xspace}

\newcommand*{\tikzmk}[1]{\tikz[remember picture,overlay,] \node (#1) {};\ignorespaces}
\newcommand{\boxit}[1]{\tikz[remember picture,overlay]{\node[xshift=-5pt,yshift=3pt,fill=#1,opacity=.25,fit={(A)($(B)+(.97\linewidth,.8\baselineskip)$)}] {};}\ignorespaces}
\newcommand{\boxitplus}[1]{\tikz[remember picture,overlay]{\node[xshift=-5pt,yshift=3pt,fill=#1,opacity=.25,fit={(A)($(B)+(.97\linewidth,-.3\baselineskip)$)}] {};}\ignorespaces}
\colorlet{red}{red!40}
\colorlet{yellow}{yellow!60}
\colorlet{blue}{cyan!60}

%% file: intro.tex
\section{Introduction}
\label{sec:intro}

Many networks use distributed routing protocols (e.g., BGP and OSPF) to compute forwarding rules. The computations are influenced by the network topology, link failure states, protocol-specific algorithms, and numerous inputs specified in router config\-urations---e.g., neighbor relationships, prefixes to advertise, link costs, route filters, etc. To ensure the network computes rules that satisfy forwarding requirements---e.g., reachability, waypoint traversals, path preferences---suitable inputs must be specified in router configurations.

Unfortunately, providing the correct inputs is complex,
and configuration errors are common~\cite{bgpmisconfig, atpg}.
This has motivated researchers to develop 
numerous systems to detect~\cite{rcc, minerals, hsa, veriflow, netplumber, anteater, minesweeper, arc, batfish, bagpipe, ynot, diffprov, tiramisu, rcdc, plankton, deltanet, selfstarter, shapeshifter, apkeep}, remediate~\cite{cpr, jinjing}, and avoid~\cite{propane, zeppelin, netcomplete, synet} configuration errors.

Error detection tools fall into four major categories. Configuration checkers~\cite{minerals, rcc, selfstarter} inspect configurations for deviations from common practice; however these tools ignore the network's forwarding requirements. Data plane verifiers~\cite{hsa, veriflow, netplumber, anteater, deltanet} analyze forwarding rule snapshots to detect violations of forwarding requirements, but they ignore the network's configurations. Control plane verifiers~\cite{minesweeper, arc, batfish, bagpipe, tiramisu, plankton, rcdc, shapeshifter} derive from the configurations a model that encodes {\em all} possible forwarding rules, but
only one control plane verifier (Batfish~\cite{batfish}) links problematic forwarding rules with the configuration statements that caused them, and this verifier scales poorly~\cite{arc}. Finally, provenance-based diagnosis tools~\cite{ynot, diffprov} provide explanations for the presence (or absence) of specific forwarding rules, but these explanations are based on observed events---e.g, route advertisements and link failures---rather than configurations. In summary, 
existing error detection tools are ill suited for identifying {\em which aspects of a network's configurations caused a forwarding requirement to be violated}. Thus, it is hard for engineers, or even automated repair tools~\cite{cpr}, to correct the error(s).

Configuration synthesizers~\cite{propane, zeppelin, netcomplete, synet, cpr, jinjing} are designed to automatically create error-free configurations, supposedly avoiding the need to detect or remediate errors. However, existing synthesizers do not cover the full range of requirements networks may need to satisfy---e.g., preventing communication between specific subnets---or the range of features required to satisfy those requirements---e.g., packet filters. Consequently, portions of a network's configurations are often still written by-hand, introducing the possibility for errors to occur. Furthermore, to make synthesis tractable, some tools require engineers to provide configuration templates/hints, which may contain errors or be overly restrictive---e.g., we uncovered an error in the templates used by NetComplete~\cite{netcomplete} to syntheisze WAN configurations (\secref{sec:evaluation:real_errors}). 

In this paper, we introduce a novel {\em configuration error localizer} (\name) that {\em precisely identifies which configuration segments contribute to the violation of one or more forwarding requirements}. 
\name constructs a system of satisfiability modulo theories (SMT) constraints that encode a network's configurations, control plane logic, and forwarding requirements,
and then checks if the constraints are satisfiable under all failure scenarios of interest. If the constraints are unsatisfiable under some failure scenario---indicating the forwarding requirements cannot always be satisfied by the current configurations---then \name computes a {\em minimal correction set}---i.e., a subset of constraints whose removal from the problem would allow the remaining constraints to be satisfied. The constraints contained in the correction set are used to flag configuration statements that contribute to the violation of forwarding requirements. 

Accurately identifying configuration errors using correction sets 
requires addressing several challenges: (1) Existing SMT-based network models~\cite{minesweeper, netcomplete, cpr} use constraints that co-mingle configuration and control logic, which makes it difficult to 
separate configuration errors from software bugs. (2) Forwarding requirement violations may be caused by {\em missing} configuration statements (e.g., a missing filter rule), but most models only encode the configuration statements that are present; only a few models encode {\em plausible} configuration statements, based on manually-specified templates~\cite{netcomplete} or a limited set of protocol options~\cite{cpr}. (3) A forwarding requirement may be satisfied under some conditions (e.g., no link failures) and violated under others.
(4) There may be multiple plausible explanations for a forwarding requirement violation. (5) Configurations may contain multiple errors whose impact varies across failure scenarios and forwarding requirements.

To overcome these challenges, \name uses a carefully structured SMT formulation and domain-specific algorithms for exploring the space of correction sets. In particular, \name separates the encoding of configuration from the encoding of control logic and explicitly encodes the absence of certain configuration 
statements---namely, those present elsewhere in the configurations.
Then, \name conducts a counterexample-guided exploration of the failure scenarios under which a forwarding requirement is violated
and employs a domain-specific version of a state-of-the-art minimal correction set (MCS) enumeration algorithm~\cite{marco} to efficiently compute multiple, configuration-focused MCSes. 
Finally, \name ranks MCSes based on common network management practices~\cite{mpa, taleoftwocampuses} to identify the most-likely configuration errors.

We implement \name atop the Minesweeper network verifier~\cite{minesweeper},
and we evaluate \name using a combination of real and synthetic network configurations. Using \name, we pinpoint several configuration errors in two real university networks as well as synthesized WAN configurations. Additionally, we show that \name identifies all routing-related and at least half of all ACL-related errors we introduce, and achieves perfect precision for 40\% of the scenarios. Finally, we show that \name localizes errors in less than 15 seconds for half of the scenarios, and less than 10 minutes in the remaining scenarios, whereas network engineers took more than a hour in over half of cases they reported.

%% file: techniques.tex
\section{Existing Techniques}
\label{sec:techniques}

To better understand the gap in approaches for localizing configuration errors,
we study the current techniques used by network engineers (\secref{sec:techniques:survey}) and review existing systems developed by researchers (\secref{sec:techniques:research}).

\input{techniques_survey}
\input{techniques_research}

%% file: techniques_survey.tex
\subsection{Resources used by network engineers}
\label{sec:techniques:survey}

To better understand how network engineers currently identify configuration errors, we surveyed \surveycount  engineers.\footnote{We advertised our survey to the North American Network Operators Group (NANOG) and EDUCAUSE Network Management Constituent Group.} We asked the engineers to consider a routing-related configuration error they recently corrected (excluding errors related to software bugs) and answer 9 questions about the error.

The engineers reported on configuration errors in enterprise campus (48\%), service provider (32\%), data center (16\%), and Internet exchange point (4\%) networks. The reported errors were approximately evenly divided between networks with less than 10 routers, 10 to 50 routers, and more than 50 routers. Almost half of the reported errors involved interface configurations. Engineers also reported on errors involving BGP, OSPF, IS-IS, MPLS, static routes, VPN, route filters, ACLs, and VLANs. Interestingly, one-third of the reported errors occurred in configuration segments that were automatically generated from templates or intents. This suggests that locating configuration errors is an important problem even with recent advances in configuration synthesis~\cite{propane, netcomplete, zeppelin, synet}.

\begin{figure}
    \centering
    \begin{subfigure}{0.48\columnwidth}
        \includegraphics[width=1\columnwidth]{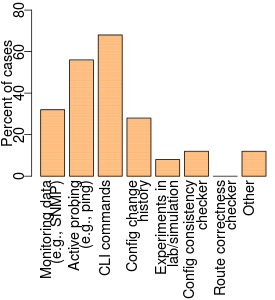}
        \caption{Used}
        \label{fig:identify_resources}
    \end{subfigure}
    \begin{subfigure}{0.48\columnwidth}
        \includegraphics[width=1\columnwidth]{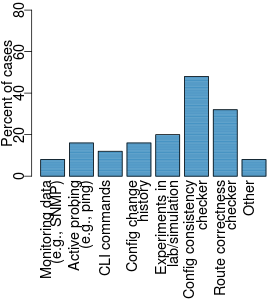}
        \caption{Desired}
        \label{fig:identify_easier}
    \end{subfigure}
    \caption{Resources for identifying errors}
\end{figure}

Engineers reported using a wide range of resources to identify configuration errors (\figref{fig:identify_resources}). Active probing (e.g., ping or traceroute) and router commands (e.g., {\tt show route}) were the most commonly used resources, while tools that check for inconsistent/missing routes were not used by any engineers. Engineers used multiple resources in 72\% of the cases.
The most frequently used combination of resources was active probing and router commands (48\% of cases). 

Engineers were able to identify and correct errors reasonably quickly: almost half of the errors were identified in less than an hour, and 90\% were identified in less than 12 hours. 

Lastly, we asked what resources would have made it easier to identify the error (\figref{fig:identify_easier}). Half of the engineers desired a tool that detects inconsistencies within/across device configurations, and one-third wanted a tool that checks for inconsistent/missing routes. Engineers also wanted tools that can detect missing route advertisements.
Below, we discuss several existing tools that fit one of these categories.

%% file: techniques_research.tex
\subsection{Tools developed by researchers}
\label{sec:techniques:research}

\begin{figure}
    \centering
    \includegraphics[width=1\columnwidth]{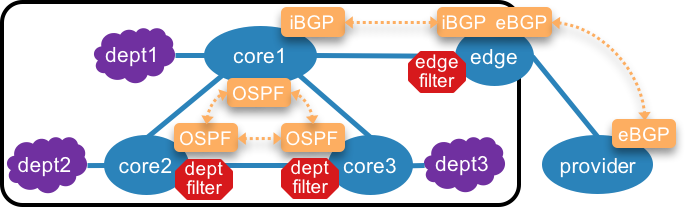}
    \caption{Example campus network: blue ovals are routers; purple clouds are subnets; solid blue lines are  links; orange rectangles are routing processes; dashed orange lines are routing adjacencies; red octagons are ACLs}
    \label{fig:techniques_example}
\end{figure}

Researchers have developed numerous tools for determining whether a network's  configurations are correct and synthesizing correct (updates to) configurations. In this section, we review five different categories of tools and illustrate the capabilities and limitations of the tools in each category using a small example network (\figref{fig:techniques_example}). 

\newcommand{\edgeRouter}{\texttt{edge}\xspace}
\newcommand{\providerRouter}{\texttt{provider}\xspace}
\newcommand{\coreOneRouter}{\texttt{core1}\xspace}
\newcommand{\coreTwoRouter}{\texttt{core2}\xspace}
\newcommand{\coreThreeRouter}{\texttt{core3}\xspace}
\newcommand{\deptFilter}{\texttt{deptFilter}\xspace}
\newcommand{\edgeFilter}{\texttt{edgeFilter}\xspace}
\newcommand{\deptOneSubnet}{\texttt{dept1}\xspace}
\newcommand{\deptTwoSubnet}{\texttt{dept2}\xspace}
\newcommand{\deptThreeSubnet}{\texttt{dept3}\xspace}
\newcommand{\deptOneTwoSubnets}{\texttt{dept1/2}\xspace}
\newcommand{\deptOneThreeSubnets}{\texttt{dept1/3}\xspace}
\newcommand{\deptTwoThreeSubnets}{\texttt{dept2/3}\xspace}
\newcommand{\externalSubnet}{\texttt{external}\xspace}
\newcommand{\deptAllSubnet}{\texttt{dept1/2/3}\xspace}
\newcommand{\reqInternal}{\texttt{FR2}\xspace}
\newcommand{\reqExternal}{\texttt{FR1}\xspace}

The example network uses the Border Gateway Protocol (BGP),  Open Shortest Path First (OSPF), and access control lists (ACLs). \edgeRouter receives routes from \providerRouter using eBGP and forwards routes to \coreOneRouter using iBGP. \coreOneRouter originates BGP routes for \deptAllSubnet. The BGP configuration on \edgeRouter, \providerRouter, and \coreOneRouter is bare-bones and does not contain any policies that filter or modify route advertisements. Routers \coreOneRouter, \coreTwoRouter, and \coreThreeRouter run OSPF. \coreOneRouter advertises a default route to \coreTwoRouter and \coreThreeRouter, because external advertisements are not propagated to these two routers. The \edgeFilter ACL defined on \edgeRouter only allows traffic to \coreOneRouter that is destined for \deptOneSubnet or \deptTwoSubnet. The \deptFilter ACL defined on \coreThreeRouter (\coreTwoRouter) only allows traffic to \coreTwoRouter (\coreThreeRouter) that originated from \deptTwoThreeSubnets.

\newcommand{\iBGPError}{\texttt{CE1}\xspace}
\newcommand{\edgeFilterError}{\texttt{CE2}\xspace}
\newcommand{\deptFilterError}{\texttt{CE3}\xspace}

The network has two forwarding requirements: (\reqExternal) all departments can access the Internet (in the absence of failures), and (\reqInternal) all departments can communicate, even in the presence of a single link failure.
However, these requirements are violated due to three different configuration errors: 

\begin{itemize}
    \setlength{\topsep}{0pt}
    \setlength{\itemsep}{0em}
    \setlength{\parskip}{0pt}
    \setlength{\parsep}{0pt}
    \item[\iBGPError:] \edgeRouter's iBGP configuration is missing an option that causes a router to set itself as the next hop in routes forwarded to iBGP peers. Consequently, the routes \edgeRouter forwards to \coreOneRouter will contain \providerRouter as the next hop, which is not directly reachable from \coreOneRouter. This causes \reqExternal to be violated.
    \item[\edgeFilterError:] \edgeFilter does not allow traffic destined for \deptThreeSubnet to be sent to \coreOneRouter. Consequently, external hosts cannot reach \deptThreeSubnet, which violates \reqExternal.
    \item[\deptFilterError:] \deptFilter does not allow traffic that originated from \deptOneSubnet. Thus, if the \coreOneRouter--\coreTwoRouter link fails, OSPF re-routes traffic from \deptOneSubnet to \deptTwoSubnet through \coreThreeRouter, and the traffic is dropped by \coreThreeRouter, which violates \reqInternal. A similar issue arises when the \coreOneRouter--\coreThreeRouter link fails.
    \vspace{-0.5em}
\end{itemize}



\minisectionpreamble{Configuration checkers}~\cite{rcc, minerals, selfstarter} statically analyze router configurations to detect anomalies. In particular, rcc~\cite{rcc} checks whether BGP configurations conform to common best practice---e.g., private addresses are not advertised externally---while Minerals~\cite{minerals} and SelfStarter~\cite{selfstarter} check for inconsistencies between routers---e.g., a filter is defined on multiple routers but contains different rules on some routers.

Configuration checkers (specifically, rcc) can only detect \iBGPError.
No checker can detect \edgeFilterError or \deptFilterError. From the perspective of Minerals and ShapeShifter, the \deptFilter ACLs on \coreTwoRouter and \coreThreeRouter contain the same rule and are applied in a consistent manner---namely, \coreTwoRouter filters incoming traffic from \coreThreeRouter and vice versa---so no problem exists with these ACLs. The \edgeFilter ACL on \edgeRouter is unique---it has a unique name, unique rules, and is applied in a unique context---so there is no point of comparison that can be used to check if this ACL contains errors. ACLs are not even considered by rcc, which is designed explicitly for checking BGP-related configuration.

\minisectionpreamble{Data plane verifiers}
\cite{hsa, netplumber, anteater, veriflow, deltanet} 
transform a snapshot of routers' forwarding information bases (FIBs) into algebraic functions~\cite{hsa}, boolean constraints~\cite{anteater}, or forwarding graphs~\cite{netplumber, veriflow, deltanet, apkeep} and use algebraic analysis, satisfiability (SAT) solvers, or graph algorithms, respectively, to detect violations of engineer-specified forwarding requirements.

Data plane verifiers can immediately detect that \reqExternal is violated, because no entries for external networks will be installed in \coreOneRouter's FIB (due to \iBGPError), and
\edgeRouter's FIB will contain a rule to discard traffic destined for \deptThreeSubnet\footnote{Routers store filtering rules separately from forwarding rules, but for simplicity we refer to them collectively as the FIB.}  (due to \edgeFilterError).
However, data plane verifiers will not detect a violation of \reqInternal (due to \deptFilterError) until the \coreOneRouter--\coreTwoRouter or \coreOneRouter--\coreThreeRouter link fails and the routers update their FIBs. 

Although data plane verifiers can detect violations of the forwarding requirements, the verifiers do not identify which configuration segments led to the missing and erroneous FIB entries. FIBs are much easier to model than router configurations and control logic, so data plane verifiers only consider the former. Identifying the cause of problematic FIB entries is left as a manual exercise for network engineers. This diagnosis is further complicated by the fact that errors in FIBs may be the result of errors in configurations and/or bugs in the control plane software running on the routers. Determining which component to blame requires additional sleuthing.

\minisectionpreamble{Control plane verifiers}
\cite{batfish, era, arc, minesweeper, bagpipe, tiramisu, plankton} model the route advertisement and selection algorithms used by various protocols (e.g., BGP and OSPF) in order to compute routers' expected FIBs under various failure scenarios. The models---e.g., Datalog programs~\cite{batfish}, 
weighted-directed graphs~\cite{arc, tiramisu}, or systems of constraints~\cite{minesweeper, bagpipe}---are based on router configurations, protocol standards, and vendor documentation. 
The models are used to check whether forwarding requirements are satisfied under all failure scenarios of interest.

Control plane verifiers can detect all three configuration errors,
and produce counterexamples with 
forwarding paths that demonstrate the forwarding requirement violations. However, akin to data plane verifiers, most control plane verifiers do not identify which segments of the configuration contributed to the problematic forwarding paths.
Only one control plane verifier (Batfish~\cite{batfish}) is capable of identifying which configuration statements led to the problematic forwarding rules, and this tool scales poorly when exploring multiple failure scenarios of interest~\cite{arc}.

\minisectionpreamble{Provenance-based diagnosis tools}~\cite{ynot, diffprov} focus on explaining {\em why} a FIB entry is present (or absent). However, these explanations are based on observed events---e.g., route advertisements and link failures---rather than routers' configurations. Consequently, the explanations are valuable for identifying control plane software bugs that lead to forwarding requirement violations, but the explanations do not help network engineers identify which configuration segments are responsible for forwarding requirement violations. 

\minisectionpreamble{Configuration synthesizers}~\cite{propane, zeppelin, netcomplete, synet, cpr, jinjing} are designed to automatically create error-free configurations, supposedly avoiding the need to localize errors. However, portions of configurations are often still written by-hand~\cite{mpa}, introducing the possibility for errors to occur. Furthermore, some synthesizers require engineers to provide configuration templates, which may contain errors---e.g., we uncovered an error in the templates provided with NetComplete~\cite{netcomplete} (\secref{sec:evaluation:real_errors}).

In summary, existing tools either {\em identify a limited set of configuration errors} or {\em fail to identify which configuration segments cause forwarding requirement violations}.

%% file: approach.tex
\section{Correction Sets}
\label{sec:approach}


Given the aforementioned limitations of existing tools, we seek to develop a framework for accurately and efficiently identifying configuration segments that contribute to forwarding requirement violations.
We view this task as a problem of identifying {\em incompatibilities} between router configurations and forwarding requirements. For example, in the network presented in \secref{sec:techniques:research}, \coreTwoRouter's configuration contains an ACL that only permits traffic destined for \deptTwoThreeSubnets, whereas one of the forwarding requirements (\reqInternal) is that all departments can communicate. This configuration segment and forwarding requirement are incompatible. Of course, incompatibilities are not always this obvious: e.g., the error in \edgeRouter's iBGP configuration (\iBGPError) is subtle.


\newcommand{\configurationConstraints}{\ensuremath{C}}
\newcommand{\controlLogicConstraints}{\ensuremath{L}}
\newcommand{\requirementConstraints}{\ensuremath{R}}

Our key insight is to identify incompatibilities 
using minimal correction sets (MCSes) derived from 
an SMT-based encoding of a network's configurations and forwarding requirements.
Several prior works~\cite{minesweeper,netcomplete,synet} have introduced SMT-based models that encode a network's configurations and control logic as a conjunction of logical constraints, $\configurationConstraints \land \controlLogicConstraints$. 
These constraints define the paths the network will compute for specific prefixes under various failures. Forwarding requirements of interest are appended as a conjunction of additional logical constraints, $\requirementConstraints$. The satisfiability of these models can be checked using an SMT solver (e.g., Z3~\cite{z3}) in order to determine whether the configurations (can be modified to) satisfy all forwarding requirements.

\begin{figure}[t]
    \centering
    \begin{subfigure}{0.75\columnwidth}
        \includegraphics[width=1\columnwidth]{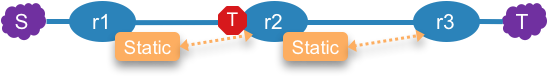}
        \caption{With ACL and static routes}
        \label{fig:challenges_example0}
    \end{subfigure}
    \begin{subfigure}{0.75\columnwidth}
        \includegraphics[width=1\columnwidth]{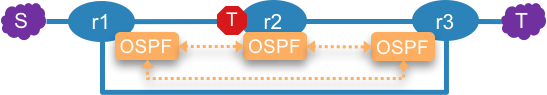}
        \caption{With ACL, OSPF, and additional link}
        \label{fig:challenges_example1}
    \end{subfigure}
    \begin{subfigure}{0.75\columnwidth}
        \includegraphics[width=1\columnwidth]{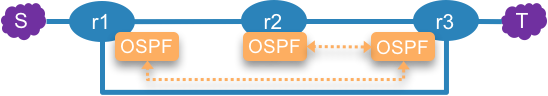}
        \caption{With missing OSPF adjacency}
        \label{fig:challenges_example2}
    \end{subfigure}
    \caption{Example networks}
    \label{fig:challenges_example}
\end{figure}

\newcommand{\routerOne}{\ensuremath{r1}}
\newcommand{\routerTwo}{\ensuremath{r2}}
\newcommand{\routerThree}{\ensuremath{r3}}
\newcommand{\subnetS}{\ensuremath{S}}
\newcommand{\subnetT}{\ensuremath{T}}
\newcommand{\forward}[3]{\ensuremath{fwd^{#1}_{#2\rightarrow#3}}}
\newcommand{\reach}[2]{\ensuremath{reach^{#1}_{#2}}}
\newcommand{\permit}[3]{\ensuremath{allow^{#1}_{#2\rightarrow#3}}}
\newcommand{\route}[3]{\ensuremath{route^{#1}_{#2\rightarrow#3}}}
\newcommand{\connected}[2]{\ensuremath{connected^{#1}_{#2}}}
\newcommand{\static}[3]{\ensuremath{static^{#1}_{#2\rightarrow#3}}}
\newcommand{\nexthop}{\ensuremath{h}}

\input{algorithms/challenges_encoding0.tex}

For example, \figref{fig:challenges_encoding0} contains a simple, partial encoding of the forwarding requirement, control logic, and router configurations for the example network in \figref{fig:challenges_example0}. (The detailed encoding used by \name is discussed in \secref{sec:encoding}.) To satisfy the requirement that subnet $\subnetT$ is reachable from $\subnetS$ (line 1), $\subnetT$ must be reachable from $\routerOne$ (line 2). $\subnetT$ is reachable from a router (e.g., $\routerOne$) iff the router either forwards directly to $\subnetT$ or to a neighboring router (e.g., $\routerTwo$) which can reach $\subnetT$ (e.g., line 3).
According to the control logic, a router (e.g., $\routerOne$) forwards traffic directly to a subnet (e.g., $\subnetT$)
iff the subnet is directly connected to the router and the traffic destined for the subnet is allowed by an ACL (e.g., line 4). Furthermore, a router (e.g., $\routerOne$) forwards traffic destined for a subnet (e.g., $\subnetT$) to another router (e.g., $\routerTwo$) if a static route is configured, the subnet is not directly connected, and traffic destined for the subnet is allowed by an ACL (e.g., line 5). 
Finally, the encoding indicates which static routes, connected subnets, and ACLs are configured (e.g., lines 6--9). Checking these constraints using an SMT solver reveals they are unsatisfiable---i.e., the forwarding requirement is violated. 



When the constraints are unsatisfiable, 
we can compute a {\em minimal correction set} (MCS)---i.e., a subset of logical constraints whose removal from the problem causes the remaining constraints to become satisfiable. For example, if we remove the constraint that traffic for $\subnetT$ is not allowed to be sent from $\routerOne$ to $\routerTwo$ (line 9 in \figref{fig:challenges_encoding0}), then the remaining constraints, including the forwarding requirement (line 1), are satisfiable. MCSes thus provide a way to identify incompatibilities between configuration segments (e.g., the ACL on $\routerTwo$ in \figref{fig:challenges_example0}) and forwarding requirements (e.g., \reqInternal).



However, accurately identifying configuration errors using MCSes requires addressing several challenges, which we articulate below.
\secsref{sec:encoding}{sec:mcses} discuss how \name addresses them.

\subsection{Challenges}
\label{sec:approach:challenges}




\minisection{Challenge 1: limiting error localization to configurations} The first challenge arises from the fact that both configurations and control logic impact whether a forwarding requirement is satisfied. 
For example, FIB entries
partially depend on: ($i$) a router's logic for populating the global routing information base (RIB) from the RIBs of individual routing protocols, and ($ii$) ACLs defined in the router's configuration. Consequently, we may attribute forwarding requirement violations to a router's control logic or its configuration.
For example, we showed above that the constraint representing the ACL on $\routerTwo$ 
(line 9 in \figref{fig:challenges_encoding0}) is an MCS---i.e., removing this constraint allows all other constraints, including the forwarding requirement (line 1), to be satisfied. However, the constraint that $\routerOne$ forwards directly to $\subnetT$ iff $\subnetT$ is directly connected (line 4) is also an MCS. The fundamental difference between these MCSes is that the former represents configuration while the latter represents control logic.

While configuration errors and software bugs are both common causes of network outages~\cite{netcraft, atpg}, we focus on localizing the former, because: ($i$) configurations are fully under network engineers' control, whereas router software is often closed-source; and ($ii$) software fault localization has already been extensively studied~\cite{faultloc_survey}. Limiting our localization to configuration errors
requires an encoding that separates configuration and control logic into separate constraints (\secref{sec:encoding:configuration}) and an MCS extraction algorithm that only produces MCSes with configuration-related constraints (\secref{sec:mcses:compute}).


\minisection{Challenge 2: partially violated forwarding requirements}
The second challenge arises from the fact that link and node failures may also impact whether a forwarding requirement is satisfied. For example, the network in \figref{fig:challenges_example1} has two loop-free physical paths from $\subnetS$ to $\subnetT$. OSPF will choose the shorter path ($\routerOne\rightarrow\routerThree$) when the $\routerOne$--$\routerThree$ link is active and the longer path ($\routerOne\rightarrow\routerTwo\rightarrow\routerThree$) when the link fails. However, $\subnetT$ is only reachable from $\subnetS$ when the shorter path is used, because an ACL on $\routerTwo$ blocks traffic destined for $\subnetT$. 

Using MCS-based error localization in this scenario is difficult due to the way existing SMT-based models handle failures. SMT-based models designed for verification~\cite{minesweeper} identify a single failure scenario (if any) under which a requirement is violated. In particular, the models include free variables representing link states and constraints encoding the negation of the forwarding requirement. The problem is satisfiable iff there exists a failure scenario under which the requirement is violated (e.g., the $\routerOne$--$\routerThree$ link fails), and unsatisfiable iff the requirement is fulfilled under every failure scenario (which is not the case for our example network). However, by definition, a satisfiable problem has no MCSes~\cite{flint}, so we cannot employ MCS-based error localization with this encoding of failures and forwarding requirements.
If we remove the negation from the forwarding requirement, then the problem is satisfiable iff there exists a  scenario under which the requirement is fulfilled (e.g., no link failures), and unsatisfiable iff the requirement is violated under every failure scenario (which is not the case for our example). So again there are no MCSes.

A different problem arises if we use an SMT-based model designed for synthesis~\cite{netcomplete}. These models avoiding introducing free variables to model link failure state by instead requiring backup (and primary) paths to be specified in the forwarding requirement. For example, we would need to reformulate our forwarding requirement to say that traffic from $\subnetS$ to $\subnetT$ is forwarded along any of ${\routerOne\rightarrow\routerThree, \routerOne\rightarrow\routerTwo\rightarrow\routerThree}$. While this is feasible to do for a small network, computing such paths for a larger network is difficult~\cite{zeppelin}.

To identify problematic configuration segments when forwarding requirements are violated under some, but not all, failure scenarios, we must employ a more sophisticated approach for considering failures (\secref{sec:mcses:failures}).

\minisection{Challenge 3: identifying missing configuration statements} While the aforementioned requirement violations were caused by the presence of an ACL, the {\em omission} of a configuration segment can also cause violations. For example, every router in \figref{fig:challenges_example2} runs OSPF and is physically connected to every other router, but OSPF is not configured on the interfaces that connect $\routerOne$ and $\routerTwo$. Consequently, if the $\routerOne$--$\routerThree$ link fails, $\subnetT$ will be unreachable from $\subnetS$. We expect such omissions occur in practice due to the lack of (full) automation in some networks (\secref{sec:techniques:survey}) and the large scope of some configuration updates~\cite{mpa}.

Existing SMT-based models~\cite{minesweeper, netcomplete} are unsuitable for locating the configuration error, because they only encode statements that are present in the configurations.\footnote{NetComplete, which is designed for configuration synthesis, accepts configuration templates containing holes, but only the holes that are explicitly present in the template are encoded in the model.} No constraints are created for absent configuration, so it is impossible to produce an MCS containing a constraint related to the configuration error. To ensure \name can detect missing configuration segments that cause requirement violations, we must explicitly encode the absence of configuration (\secref{sec:encoding:absent}).


\minisection{Challenge 4: ranking plausible errors} There are often many ways to satisfy a forwarding requirement. Thus, if a requirement is violated, there are often many MCSes: e.g., for 75\% (45\%) of the configuration errors we introduce into real university networks (\secref{sec:evaluation:accuracy}), there are more than 20 (100) MCSes.

Requiring network engineers to sift through a large list of plausible errors places an undue burden on the engineers and undermines the motivation for \name. Consequently, it is essential to rank errors based on their likelihood of being true errors. However, what constitutes a ``true error'' depends on engineers' network management practices~\cite{mpa}. To rank configuration error candidates, we must consider what design and operational practices a network follows (\secref{sec:mcses:rank}).

\minisection{Challenge 5: identifying multiple configuration errors} The above examples only contain one configuration error, but real configurations often contain multiple errors~\cite{selfstarter, minerals, batfish, era}. In the simplest case the errors are related, insofar as they impact the same set of requirements under the same set of failures. For example, if we added an ACL on $\routerTwo$ in the network in \figref{fig:challenges_example2}, then both the ACL and the absence of an OSPF relationship between $\routerOne$ and $\routerTwo$ would cause $\subnetT$ to be unreachable from $\subnetS$ when the $\routerOne$--$\routerThree$ link fails. 
However, it is also possible for different requirement violations to be caused by different errors. For example, assume we instead configured an ACL on $\routerOne$'s interface to $\routerThree$. When the $\routerOne$--$\routerThree$ link is active, the ACL causes $\subnetT$ to be unreachable from $\subnetS$; when the $\routerOne$--$\routerThree$ link fails, the absence of an OSPF relationship between $\routerOne$ and $\routerTwo$ causes $\subnetT$ to be unreachable. Ensuring all errors are identified requires a carefully designed MCS enumeration algorithm (\secref{sec:mcses})

\secsref{sec:encoding}{sec:mcses} discuss how \name addresses these challenges.

%% file: algorithms/challenges_encoding0.tex
\begin{figure}[t]
    \begin{algorithm}[H]
    \small
    \tikzmk{A}
    $\reach{\subnetT}{\subnetS}$\;
    $\reach{\subnetT}{\subnetS}\iff\reach{\subnetT}{\routerOne}$\;
    $\reach{\subnetT}{\routerOne}\iff
        \forward{\subnetT}{\routerOne}{\subnetT}$
        $\lor (\forward{\subnetT}{\routerOne}{\routerTwo} 
            \land \reach{\subnetT}{\routerTwo})$\;
    \tikzmk{B}\boxit{red}
    \tikzmk{A}
    $\forward{\subnetT}{\routerOne}{\subnetT}\iff
        \connected{\subnetT}{\routerOne}
        \land \permit{\subnetT}{\routerOne}{\subnetT}$\;
    $\forward{\subnetT}{\routerOne}{\routerTwo}\iff
        \lnot\connected{\subnetT}{\routerOne}
        \land\static{\subnetT}{\routerOne}{\routerTwo}
        \land \permit{\subnetT}{\routerOne}{\routerTwo}$\;
    \tikzmk{B}\boxit{yellow}
    \tikzmk{A}
    $\static{\subnetT}{\routerOne}{\routerTwo}$\;
    $\static{\subnetT}{\routerTwo}{\routerThree}$\;
    $\connected{\subnetT}{\routerThree}$\;\tikzmk{B}\boxitplus{blue}
    $\lnot\permit{\subnetT}{\routerOne}{\routerTwo}$\;
    \end{algorithm}
    \caption{Partial encoding of the forwarding requirement (red), control logic (yellow), and configuration (blue) for the example network in \figref{fig:challenges_example0}}
    \label{fig:challenges_encoding0}
\end{figure}

%% file: encoding.tex
\section{SMT-based Network Encoding}
\label{sec:encoding}


As discussed in \secref{sec:approach:challenges}, \name requires an SMT-based network model that satisfies two important properties: 
($i$) configuration is encoded separately from control logic, so we can limit our localization to configuration errors; and ($ii$) the absence of configuration is explicitly encoded, so we can identify omissions that contribute to forwarding requirement violations.  Unfortunately, none of the existing SMT-based network models designed for control plane verification/synthesis~\cite{minesweeper, netcomplete, bagpipe, jinjing} satisfy these properties. However, Minesweeper~\cite{minesweeper} and NetComplete~\cite{netcomplete} accommodate a broad range of configuration constructs,
so we use these models as a starting point, and adapt them to satisfy the aforementioned requirements.


\input{encoding_background}
\input{encoding_configuration}
\input{encoding_absent}

%% file: encoding_background.tex
\subsection{Existing Models}
\label{sec:encoding:background}

Minesweeper and NetComplete encode a network's forwarding behavior as a system of logical constraints over symbolic route advertisements. Symbolic advertisements are created for each routing adjacency---i.e., pair of neighboring routers that are configured to exchange routes using a particular routing protocol. Each symbolic advertisement is composed of multiple symbolic variables that mirror the fields in actual route advertisements: e.g., prefix, path cost/length, and local preference. Constraints on these symbolic advertisements express: ($i$) the route advertisement import/export behavior of each router---which includes the application of route policies that filter and/or modify route advertisements, the forwarding of route advertisements, and the origination of route advertisements; ($ii$) the protocol-specific and cross-protocol route selection procedures within each router; and ($iii$) the consequent packet forwarding behavior. We explain these constraints in detail in \appref{sec:encoding:appendix}.

%% file: encoding_configuration.tex
\subsection{Encoding Configuration}
\label{sec:encoding:configuration}

As discussed earlier, configuration must be encoded separately from control logic in order to limit \name's output to configuration errors. However, the constraints generated by Minesweeper and NetComplete co-mingle configuration and control logic. For example, the constraint that encodes the routes \coreOneRouter exports to \coreTwoRouter (\figref{fig:encoding_core1_export} in \appref{sec:encoding:appendix}) includes the OSPF originated prefix for \deptOneSubnet ({\tt 1.0.1.0/24}) and link costs specified in \coreOneRouter's configuration. Similarly, the constraint that encodes which packets \coreTwoRouter forwards to \coreThreeRouter (\figref{fig:encoding_core2_forward} in \appref{sec:encoding:appendix}) includes the \deptFilter ACL in \coreTwoRouter's configuration.

\name separates configuration from control logic by: ($i$) replacing configuration-based expressions in the aforementioned route import/export, route selection, and packet forwarding constraints with symbolic configuration variables; and ($ii$) adding a constraint for each configuration variable that equates the variable with the expression specified in the current configurations. 
\figref{fig:encoding_symbolic} illustrates this encoding.

\input{algorithms/encoding_symbolic.tex}

Only expressions that are {\em solely based on configuration} are replaced with symbolic configuration variables. Expressions that correspond to control logic are left unchanged. For example, the prefix \coreOneRouter is configured to originate ({\tt 1.0.1.0/24}) is replaced with a symbolic variable\linebreak ($\configOspfOriginate{\coreOneRouter}{1.0.1.0/24}$), because \coreOneRouter's configuration lists each network advertised by OSPF. However, the comparison between the originated prefix and the $\destination$ prefix specified in the forwarding requirement is left in the export constraint (line 2 in \figref{fig:encoding_symbolic_core1_export}), because a router always forwards packets according to their destination IP address. Similarly, the cost \coreOneRouter adds to advertisements sent to \coreTwoRouter is replaced with a symbolic variable ($\configOspfCost{\coreOneRouter}{\coreTwoRouter}$), because \coreOneRouter's configuration lists the OSPF cost for each interface, but the arithmetic expression is left in the export constraint (line 8 in \figref{fig:encoding_symbolic_core1_export}), because OSPF costs are always additive. In contrast, the ACL-based expressions in forwarding constraints (e.g., line 2 in \figref{fig:encoding_core2_forward}) are replaced with a symbolic configuration variable ($\configOutAcl{\coreTwoRouter}{\coreThreeRouter}$), because packet filters defined in a router's configuration specify both the packet fields and values to match.

It is important to note that \name's use of symbolic configuration variables is fundamentally different from NetComplete's use of symbolic configuration variables. In particular, \name replaces {\em all} configuration-based expressions with symbolic variables, whereas NetComplete only uses symbolic configuration variables for holes in configuration templates; concrete configuration values are still embedded directly in NetComplete's route import/export constraints. Furthermore, \name fully constrains the value of symbolic configuration variables, whereas symbolic configuration variables in NetComplete are (partially) unconstrained free variables.



%% file: algorithms/encoding_symbolic.tex
\begin{figure}[t]
    
    \begin{subfigure}{1\columnwidth}
    \begin{algorithm}[H]
    \small
    \If{\configOspfAdjacency{\coreOneRouter}{\coreTwoRouter}}{
    \If{$\destination \in \configOspfOriginate{\coreOneRouter}{1.0.1.0/24}$}{
        $\outgoing{OSPF}{\coreOneRouter}{\coreTwoRouter}.\prefix = \configOspfOriginate{\coreOneRouter}{1.0.1.0/24}$\;
        $\outgoing{OSPF}{\coreOneRouter}{\coreTwoRouter}.\cost = \configOspfCost{\coreOneRouter}{\coreTwoRouter}$\;
    }
    \ElseIf{$\best{OSPF}{\coreOneRouter}.\valid \land \configOspfFilterDefault{\coreOneRouter}$}{
        $\outgoing{OSPF}{\coreOneRouter}{\coreTwoRouter}.\valid = \true$\;
        $\outgoing{OSPF}{\coreOneRouter}{\coreTwoRouter}.\prefix = \best{OSPF}{\coreOneRouter}.\prefix$\;
        $\outgoing{OSPF}{\coreOneRouter}{\coreTwoRouter}.\cost = \best{OSPF}{\coreOneRouter}.\cost + \configOspfCost{\coreOneRouter}{\coreTwoRouter}$\;
    }
    \lElse{$\outgoing{OSPF}{\coreOneRouter}{\coreTwoRouter}.\valid = \false$}
    }
    \lElse{$\outgoing{OSPF}{\coreOneRouter}{\coreTwoRouter}.\valid = \false$}
    \end{algorithm}
    \caption{\coreOneRouter export to \coreTwoRouter}
    \label{fig:encoding_symbolic_core1_export}
    \end{subfigure}
    
    \begin{subfigure}{1\columnwidth}
    \begin{algorithm}[H]
    \small
    $\forward{}{\coreTwoRouter}{\coreThreeRouter} \Leftrightarrow \best{Overall}{\coreTwoRouter} = \incoming{OSPF}{\coreTwoRouter}{\coreThreeRouter} \land\ \configOutAcl{\coreTwoRouter}{\coreThreeRouter}$
    \end{algorithm}
    \caption{\coreTwoRouter forward to \coreThreeRouter}
    \label{fig:encoding_symbolic_core2_forward}
    \end{subfigure}
    
    \begin{subfigure}{1\columnwidth}
    \begin{algorithm}[H]
    \small
    $\configOspfOriginate{\coreOneRouter}{1.0.1.0/24} = 1.0.1.0/24$\;
    $\configOspfCost{\coreOneRouter}{\coreTwoRouter} = 1$\;
    $\configAcl{\coreTwoRouter}{\deptFilter} = \source \in 1.0.2.0/24 \lor \source \in 1.0.3.0/24$\;
    $\configOutAcl{\coreTwoRouter}{\coreThreeRouter} = \configAcl{\coreTwoRouter}{\deptFilter}$\;
    $\configOutAcl{\coreTwoRouter}{\coreOneRouter} = \true$\;
    \end{algorithm}
    \caption{Configuration}
    \label{fig:encoding_symbolic_config}
    \end{subfigure}
    
    \caption{Constraints with symbolic configuration variables}
    \label{fig:encoding_symbolic}
\end{figure}

%% file: encoding_absent.tex
\subsection{Encoding Absent Configuration}
\label{sec:encoding:absent}

As illustrated above, the contents of constraints are based on the contents of routers' configurations: e.g., originated prefixes, link costs, and packets filters. However, as demonstrated in \secref{sec:approach:challenges}, the absence of configuration is also meaningful and must be encoded in the network model.

Unfortunately, the space of potentially omitted configuration is extremely large: e.g., configurations from a large ISP contain 56 different top-level commands~\cite{adaptiveparsing}, and the command reference for the latest version of Cisco IOS contains hundreds of commands~\cite{ciscoiosxecommands}. Including every possible configuration variable in the system of constraints would vastly increase the size of the problem and be prohibitively expense.

Fortunately, prior work on software fault localization has shown that erroneously omitted statements are often present (in a similar form) elsewhere in the code~\cite{codeomission}. A review of configuration omission errors reported in prior works~\cite{rcc, minerals, era} and the high prevalence of configuration clones/templates~\cite{complexitymetrics, selfstarter, minerals} suggests omissions in network configurations are also likely present (in a similar form) elsewhere in the configurations. Consequently, we use the configuration statements present in the current configurations to guide the inclusion of potentially omitted configuration variables.

In particular, \name determines which types of configuration variables (e.g., $\configOspf{originate}{}$, $\configOspf{cost}{}$, $\config{OutACL}{}$) appear in the network configurations, and includes all plausible instances of these variables in control logic (i.e., route import/export, route selection, and packet forwarding) constraints. For example, ACLs are configured on some interfaces in the example campus network in \figref{fig:challenges_example}. Thus, \name adds a $\config{OutACL}{}$ variable to every packet forwarding constraint (e.g., \figref{fig:encoding_symbolic_core2_forward}), even if no ACL is currently configured on the corresponding interface. \name also adds a configuration constraint for every $\config{OutACL}{}$ variable, which asserts the variable equals the currently configured ACL (e.g., line 4 in \figref{fig:encoding_symbolic_config}), or $\true$ (i.e., permit all traffic) if no ACL is configured for the interface (e.g., line 5). If a specific configuration option (e.g., $\configOspf{filter}{}$) is never used in the network, then no configuration variables of this type are introduced.

Two additional principles govern \name's inclusion of configuration variables for potential omissions. First, we  assume network engineers will not erroneously omit an entire routing protocol from a device's configuration. Hence, we only add omission variables for a protocol (e.g., $\configOspf{cost}{}$) if that protocol is currently enabled on the device. Second, we consider the semantics of each potentially omitted configuration option, and only introduce configuration variables for feasible instances. For example, \name only adds $\configOspf{orignate}{}$ variables for directly connected prefixes that are not already configured to be announced by OSPF; announcing other prefixes (without enabling route redistribution) is infeasible.

%% file: mcses.tex
\section{Computing MCSes}
\label{sec:mcses}

We now turn our attention to the task of computing MCSes as a means of identifying configuration errors that contribute to forwarding requirement violations. As discussed in \secref{sec:approach}, an MCS is a subset of logical constraints whose removal from the problem allows the problem to be satisfied. More formally, let $\allConstraints = 
\{\configConstraint{1},\ldots,\configConstraint{|\configConstraints|},\logicConstraint{1},\ldots,\logicConstraint{|\logicConstraints|},\requireConstraint{1},\ldots,\requireConstraint{|\requireConstraints|}\}$
represent the conjunction (or union) of the constraints
that encode a network's configurations ($\configConstraints$), control logic ($\logicConstraints$), and forwarding requirements ($\requireConstraints$). An MCS is a subset of clauses $\mcs \subseteq \allConstraints$ such that $\allConstraints \setminus \mcs$ is satisfiable and $\forall \genericConstraint{} \in \mcs : (\allConstraints \setminus \mcs) \cup \{ \genericConstraint{} \}$ is unsatisfiable. Our goal is to compute MCSes such that $\mcs \subseteq \configConstraints$,
because a router's control logic is typically outside of a network engineer's control  (\secref{sec:approach:challenges}) and a network's forwarding requirements are fixed.
Additionally, we seek to prioritize MCSes that are more likely to represent ``true errors'' (\secref{sec:approach:challenges}).

In this section, we discuss three key mechanisms we use to achieve this goal: ($i$) a counterexample-guided exploration of the failure scenarios under which a forwarding requirement is violated; ($ii$) a domain-specific MCS enumeration algorithm based on MARCO~\cite{marco}; and ($iii$) an MCS ranking algorithm inspired by engineers' network management practices~\cite{mpa}.

\input{mcses_failures}
\input{mcses_compute}
\input{mcses_rank}

%% file: mcses_failures.tex
\subsection{Handling link/node failures}
\label{sec:mcses:failures}

As discussed in \secref{sec:approach:challenges}, existing SMT-based network models~\cite{minesweeper, netcomplete} are ill-suited for localizing configuration errors that only arise under some failure scenarios.
Minesweeper's encoding of link/node failures using free variables causes $\allConstraints$ to be {\em satisfiable} (i.e., $\mcs = \varnothing$) if there is at least one failure scenario under which the forwarding requirement is (not) violated. NetComplete's encoding of primary/backup paths in the forwarding requirements scales poorly.

\begin{figure}
    \centering
    \includegraphics[width=0.7\columnwidth]{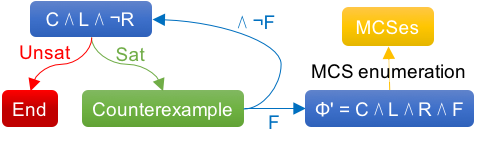}
    \caption{Counterexample-guided exploration of failures}
    \label{fig:failure_enumeration}
\end{figure}

Consequently, \name uses a {\em counterexample-guided exploration of failure scenarios} (\figref{fig:failure_enumeration}) to compute MCSes and localize errors. First, we augment each route import constraint
(e.g., \figref{fig:encoding_core1_import} in \appref{sec:encoding:appendix}) with a conditional expression representing the availability of the corresponding link (e.g., $\neg \failedLink{\coreOneRouter}{\coreTwoRouter}$).\footnote{It is trivial to extend \name to also support node failures.} Then, we check whether $\configConstraints \land \logicConstraints \land \neg \requireConstraints$ is satisfiable.
If the problem is unsatisfiable, then the forwarding requirement is never violated (i.e., $\neg \requireConstraints = \false$) and the process terminates. 
Otherwise, the solver produces a satisfying solution which represents a counterexample: i.e., a scenario under which the forwarding requirement is violated.
The satisfying solution includes concrete values for the free variables representing link failure states (e.g., $\failedLink{\coreOneRouter}{\coreTwoRouter}$). We produce an expression $\failedConstraints$ that equates each failure state variable with its concrete value: e.g., $\failedLink{\coreOneRouter}{\coreTwoRouter} = \true \land \failedLink{\coreOneRouter}{\coreThreeRouter} = \false \land \ldots$ 

We 
construct a new system of constraints:  $\allConstraints' = \configConstraints \land \logicConstraints \land  \requireConstraints \land \failedConstraints$. This system of constraints is unsatisfiable, because $\failedConstraints$ represents a failure scenario in which the forwarding requirement is violated (i.e., $\neg \requireConstraints = \true$). Hence, we can use $\allConstraints'$ to compute MCSes (\secref{sec:mcses:compute}) and localize errors that manifest under this failure scenario.

However, as demonstrated in \secref{sec:approach:challenges}, different configuration errors may manifest under different failure scenarios. To ensure we detect all of these errors, we must consider all failure states under which the forwarding requirement is violated. To achieve this, we conjunct $\neg \failedConstraints$ with $\configConstraints \land \logicConstraints \land \neg \requireConstraints$ and check if the updated problem has a satisfying solution. Any solution will contain a different combination of link failures, since the original failure scenario is disallowed by the inclusion of $\neg \failedConstraints$ in the system of constraints. We repeat this process until no more satisfying solutions (i.e., counterexamples) exist.

Note that multiple failure scenarios may result in the same forwarding behavior. For example, if we modify the network in \figref{fig:challenges_example2} to include an ACL on $\routerThree$'s incoming link from $\routerOne$, traffic from $\subnetS$ to $\subnetT$ will be forwarded along the path $\routerOne \rightarrow \routerThree$ and dropped at $\routerThree$ when any of the following sets of links fail: $\varnothing$, $\{\routerOne$--$\routerTwo\}$, $\{\routerTwo$--$\routerThree\}$, or $\{\routerOne$--$\routerTwo, \routerTwo$--$\routerThree\}$. 
As an optimization, we can formulate $\allConstraints'$ and compute MCSes using just one of these failure scenarios, since the network behaves equivalently, and hence manifests the same configuration error(s), under each of these failure scenarios. 

%% file: mcses_compute.tex
\subsection{Computing MCSes}
\label{sec:mcses:compute}

We now present our domain-specific algorithms for computing MCSes for a set of unsatisfiable network constraints $\allConstraints' = \{\configConstraint{1},\ldots,\configConstraint{|\configConstraints|},\logicConstraint{1},\ldots,\logicConstraint{|\logicConstraints|},\requireConstraint{1},\ldots,\requireConstraint{|\requireConstraints|},\failedConstraint{1},\ldots,\failedConstraint{|\failedConstraints|}\}$. 
We begin with a simple algorithm for computing a single MCS. We then use this as a subroutine of an algorithm for enumerating all MCSes. Computing multiple MCSes
is essential for exposing additional/alternative configuration errors.
Both algorithms are based on a state-of-the-art MCS enumeration algorithm called MARCO~\cite{marco}.

\minisection{Background}
Before delving into our algorithms, we highlight two additional types of constraint sets that are relevant to MCSes. A {\em maximal satisfiable subset} (MSS) is a subset of constraints $\mss \subseteq \allConstraints'$ such that $\mss$ is satisfiable and $\forall \genericConstraint{} \in \allConstraints' \setminus \mss : \mss \cup \{\genericConstraint{}\}$ is unsatisfiable. An MCS is the complement of an MSS: i.e., $ \mcs = \allConstraints' \setminus \mss$. A {\em minimal unsatisfiable subset} (MUS) is a subset of constraints $\mus \subseteq \allConstraints'$ such that $\mus$ is unsatisfiable and $\forall \genericConstraint{} \in \mus : \mus \setminus \{\genericConstraint{}\}$ is satisfiable. An MCS is a minimal hitting set of a constraint system's MUSes: i.e., an MCS contains at least one constraint from every MUS~\cite{dekleer, reiter}.

\subsubsection{Computing a single MCS}
\label{sec:mcses:compute:single}

We can compute an MCS by computing an MSS ($\mss$) and taking its complement ($\mcs = \allConstraints' \setminus \mss$).
MARCO~\cite{marco} employs a simple algorithm for computing an MSS (\figref{fig:grow_algorithm}): start with an empty set (line 1); iteratively add constraints to the set (lines 2--3); check the satisfiability of the constraint set after each addition (line 4); if adding a constraint causes the set of constraints to become unsatisfiable, then remove the most-recently-added constraint before adding the next constraint (lines 4--5).


\input{algorithms/compute_single_mcs.tex}

Unfortunately, this simple algorithm does 
not guarantee that only configuration-related constraints are included in the MCS (i.e., $\mcs \subseteq \configConstraints$).
Thus, instead of starting with $\mss = \varnothing$, we initialize $\mss$ to the set of all \mbox{logic-,} requirement-, and failure-related constraints (i.e., $\mss = \logicConstraints \cup \requireConstraints \cup \failedConstraints$).
This ensures the MSS always includes all non-configuration-related constraints (i.e., $\mss \supseteq \allConstraints' \setminus \configConstraints$), and the MCS, which is the complement of the MSS, only includes configuration-related constraints (i.e., $\mcs \subseteq \configConstraints$). This avoids pointless correction sets with constraints that cannot be modified through configuration changes. 

With this change, the time complexity for generating a single MCS for a system of constraints is linear in the number of configuration constraints. However, configuration errors are often (relatively) small~\cite{minerals, selfstarter}, so MCSes will usually be small; correspondingly, MSSes will usually include most configuration constraints. This motivates us to 
employ a divide-and-conquer approach to compute MCSes in $O(log\ n)$: instead of adding configuration constraints to $\mss$ one at a time, we split the configuration constraints in half and add either half to $\mss$. If $\mss$ becomes unsatisfiable, we recursively divide that half and try to add each of these smaller sets of configuration constraints. If $\mss$ is satisfiable after adding a group of configuration constraints, we do not recurse.  

\subsubsection{Enumerating all MCSes}
\label{sec:mcses:compute:all}

\begin{figure*}
    \centering
    \includegraphics[width=0.9\textwidth]{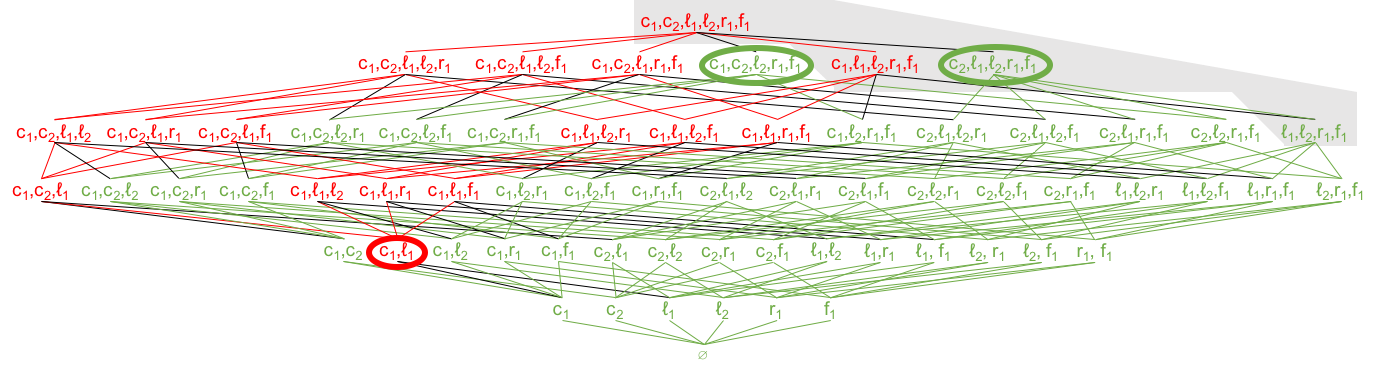}
    \vspace{-0.5em}
    \caption{Power set lattice for example network constraints; the sublattice \name explores is highlighted in gray}
    \label{fig:lattice_example}
\end{figure*}

To enumerate all MCSes, while preserving the restriction that MCSes only contain configuration-related constraints, we modify the MARCO algorithm~\cite{marco}. MARCO methodically explores the power set lattice for a set of constraints ($\allConstraints'$) to identify the frontier between the satisfiable and unsatisfiable regions. \figref{fig:lattice_example} shows the lattice for a small example set of network constraints $\allConstraints' = \{\configConstraint{1},\configConstraint{2},\logicConstraint{1},\logicConstraint{2},\requireConstraint{1},\failedConstraint{1}\}$. Every subset of $\allConstraints'$ is either satisfiable (green) or unsatisfiable (red). MSSes and MUSes are local maxima (green circles) and minima (red circles), respectively, along the frontier between the satisfiable and unsatisfiable regions. Since every subset (superset) of a (un)satisfiable set is (un)satisfiable, the set of all MSSes (MUSes) fully specify the frontier~\cite{dekleer,reiter}.

MARCO efficiently identifies the frontier using the following algorithm: select an unexplored (i.e., uncolored) subset $\genericSubset$ in the power set lattice; check the satisfiability of $\genericSubset$; if $\genericSubset$ is (un)satisfiable grow (shrink) $\genericSubset$ to an MSS (MUS) using the algorithm in \figref{fig:grow_algorithm} (or a variant that removes one constraint at a time) with $\genericSubset$ used as the initial set; mark the MSS (MUS) and all of its subsets (supersets) in the lattice as (un)satisfiable. The algorithm terminates when the satisfiability of all subsets is known (i.e., the lattice is fully colored).


However, as discussed above, we only care about MSSes that include all non-configuration related constraints (i.e., $\mss \supseteq \allConstraints' \setminus \configConstraints$). 
In other words, we only need to run MARCO on the sublattice whose bottom is $\allConstraints' \setminus \configConstraints$  and whose top is $\allConstraints'$ (e.g., the gray region in \figref{fig:lattice_example}). This has the beneficial side effect of substantially reducing the size of the region MARCO needs to explore. 

%% file: algorithms/compute_single_mcs.tex
\begin{figure}[t]
    \begin{algorithm}[H]
    \KwIn{$\allConstraints'$, a set of unsatisfiable constraints}
    $\mss = \varnothing$\;
    \For{$\genericConstraint{} \in \allConstraints'$}{
        $\mss = \mss \cup \{\genericConstraint{}\}$\;
        \If{$\mss$ is unsat}{
            $\mss = \mss \setminus \{\genericConstraint{}\}$\;
        }
    }
    \KwOut{$\mss$, an MSS}
    \end{algorithm}
    \caption{Algorithm for computing a single MSS}
    \label{fig:grow_algorithm}
\end{figure}

%% file: mcses_rank.tex
\subsection{Ranking MCSes}
\label{sec:mcses:rank}

As we discussed in \secref{sec:approach:challenges}, there are often many MCSes: e.g., for 75\% (45\%) of the synthetic configuration errors we introduce in real university networks (\secref{sec:evaluation:accuracy}), there are more than 20 (100) MCSes. The large number of MCSes stems from the multitude of ways a forwarding requirement may be satisfied~\cite{synet} and/or the presence of multiple configuration errors~\cite{selfstarter, minerals, batfish, era}. Requiring network engineers to sift through a large list of plausible errors places an undue burden on the engineers and undermines the motivation for \name. Consequently, filtering/prioritizing MCSes is essential for \name to be useful in practice.

Based on our prior observation that configuration errors are often (relatively) small~\cite{minerals, selfstarter} (\secref{sec:mcses:compute:single}), we prioritize smaller MCSes. We also aggregate MCSes across all forwarding requirement violations and failure scenarios to eliminate duplicates and ensure we cover all errors. We show in \secref{sec:evaluation:accuracy} that this approach works well in practice.


%% file: evaluation.tex
\section{Implementation \& Evaluation}
\label{sec:evaluation}

We implemented \name atop the Minesweeper network verifier in $\approx6k$ lines of Java code. Our current implementation can localize errors to specific interface states (i.e., up/down), layer-3 adjacencies, OSPF and BGP routing adjacencies, originated prefixes, OSPF link costs, route filter definitions, uses of route filters, ACL definitions, and uses of ACLs. We plan to make our implementation open source.

We evaluate \name along three dimensions: ($i$) Can \name locate real configuration errors? ($ii$) How accurate is \name's error localization? ($iii$) How quickly can \name locate errors?


\newcommand{\colgate}{{\em UnivC}\xspace}
\newcommand{\uwmadison}{{\em UnivA}\xspace}
\newcommand{\northwestern}{{\em UnivB}\xspace}

\minisection{Configurations}
We use real configurations from two university networks 
and synthetic configurations for 
eight wide area networks (WANs). 
The real configurations come from the universities' core and distribution routers.\footnote{We ignore access switches, because they primarily operate at layer 2.} \uwmadison has 11 routers and supports 65k users; \northwestern has 28 routers and supports 30k users. Both universities use OSPF for internal routes and iBGP for external routes; we exclude the latter from our analysis, because we do not know the networks' external routes.
The synthetic configurations are modeled on eight real 
WAN~\cite{topologyzoo} topologies that range in size from 34 to 159 routers.
The configurations, which use eBGP, are synthesized using NetComplete~\cite{netcomplete}. All configurations are written in Cisco's IOS language.


\minisection{Forwarding requirements}
We do not know the complete forwarding requirements for the university networks, so we only consider: ($i$) reachability between subnets associated with different departments, 
($ii$) protection of management subnets (\uwmadison only), and ($iii$) protection against source-spoofing (\northwestern only). For the WANs, we only consider reachability between pairs of edge routers.

\minisection{Experimental setup} We use a server with a 10-core 2.4 Ghz processor and 128GB of memory. Unless otherwise noted, we limit \name's MCS enumeration time to 10 minutes.

\input{evaluation_real_errors.tex}
\input{evaluation_accuracy.tex}
\input{evaluation_efficiency.tex}

%% file: evaluation_real_errors.tex
\subsection{Locating real configuration errors}
\label{sec:evaluation:real_errors}

We used 
\name to check each network's compliance with the aforementioned forwarding requirements.
Below, we highlight several types of violations we found and the corresponding configuration errors \name located.

\minisection{ACL not applied}
One of \uwmadison's forwarding requirements is to only allow certain, trusted sources to access device management VLANs. However, \name reported that this requirement was violated for two of the device management VLANs. For each of these violations, \name computed about a dozen MCSes, each of which contained configuration constraints related to VLAN interface state (up) or the absence of an ACL on some VLAN. This allowed us to quickly diagnose the problem as a missing ACL on the device management VLANs. Configuration consistency checkers~\cite{minerals, shapeshifter} cannot easily detect this error, because there are no clear outliers: some VLANs should have the ACL applied, and some VLANs should not have the ACL applied; the correct behavior for a given VLAN can only be determined from \uwmadison's forwarding requirements, which consistency checkers do not consider.

\minisection{Incorrect ACL rules}
One of \northwestern's forwarding requirements is to prevent source spoofing at VLAN granularity---i.e., a malicious host should not be allowed to impersonate a host within a different VLAN. However, \name reported that this requirement was violated for two VLANs. For each of these violations, \name computed two MCSes: one localized the error to a specific ACL definition and the other localized the error to the application of that ACL on a specific VLAN. Upon looking at the configurations, we immediately noticed that each of the identified ACLs was applied only to the VLAN for whom the forwarding requirement was violated, and the network address listed in the ACL did not match the address of the VLAN to which the ACL was applied. Although the ACL defined for each VLAN is based on a standard template, configuration consistency checkers~\cite{minerals, shapeshifter} cannot easily detect this error, because the parameter values are unique in each instantiation of this template and must match the address of the VLAN to which the ACL is applied.

\minisection{Improperly configured backup paths}
The \uwmadison network has redundant core and distribution routers and links, such that the network can tolerate three simultaneous link failures. However, \name reported that departments connected to a particular pair of primary/backup routers would be unreachable if three links failed. 
\name identified a single class of failure scenarios that caused the violation, and computed a single MCS that localized the problem to the absence of a layer-3 adjacency between a particular pair of VLAN interfaces on the backup distribution and core routers. Existing SMT-based network models~\cite{minesweeper, netcomplete} which \name is based on do not explicitly model layer 2, so we had to manually inspect the configuration based on \name's output to locate the precise configuration error. Based on \name's output, we were able to focus on configuration related to a specific VLAN on two specific routers and quickly determine that the distribution router's interface to the the core router was not configured to participate in the VLAN. In the future, we plan to extend \name to explicitly model layer 2.

\minisection{Improperly configured route filters}
The sample WAN router configuration template provided with NetComplete~\cite{netcomplete} is designed to support reachabiliby between every pair of edge routers using a fixed primary or backup path.
However, when the backup paths for different router pairs overlap, \name reported that the edge routers are unreachable if their primary links fail. For each of these violations, \name identified a single class of failure scenarios that caused the violation, and computed a single MCS that localized the problem to the match criteria used in the route filter on the overlapping router(s). Upon examining these route filters, we quickly noticed that all rules in the route filter used the same prefix list, which only contained the prefix for one of the edge routers, whereas different rules should have used different prefix lists. The authors of NetComplete accepted the bug fix we submitted. This error highlights the need for \name even in the presence of tools that (partially) generate configurations.

%% file: evaluation_accuracy.tex
\subsection{Accuracy}
\label{sec:evaluation:accuracy}

Next, we evaluate \name's accuracy. We focus on two metrics: {\em recall}---i.e., what fraction of a network's configuration errors are identified by \name; and {\em precision}---i.e., what fraction of configuration segments identified by \name contain errors. Unless otherwise noted, we only consider the smallest MCSes.

\newcommand{\OmitNeighbor}{{\em OmitNb}\xspace}
\newcommand{\OmitNetwork}{{\em OmitNw}\xspace}
\newcommand{\OmitAcl}{{\em OmitAcl}\xspace}
\newcommand{\OmitAclRule}{{\em OmitAclRule}\xspace}
\newcommand{\AddAcl}{{\em ExtraAcl}\xspace}
\newcommand{\ModifyCost}{{\em ModCost}\xspace}

\input{tables/synthetic_errors.tex}

\minisection{Errors}
Since we do not have access to a list of known errors in the real configurations and the synthetic configurations are designed to be error-free, we introduce synthetic errors into the configurations. The errors we introduce (\tabref{tbl:synthetic_errors}) are based on the real errors we found (\secref{sec:evaluation:real_errors}) as well as common errors identified in previous measurement studies~\cite{atpg, bgpmisconfig}. We generate three sets of faulty configurations per-network per-error-type, with a few exceptions: we do not introduce \OmitNetwork into \northwestern, because its OSPF stanzas contains a single {\tt network} statement covering all prefixes; and we do not introduce \OmitAcl, \OmitAclRule, or \AddAcl into the synthetic WAN configurations, because they do not use ACLs.

For each type of error, we randomly select which statements we add/remove/modify. In some instances, we remove the same/related statements from multiple devices to avoid introducing errors that could be easily detected by configuration consistency checkers~\cite{rcc, minerals, shapeshifter}: e.g., for \OmitNeighbor errors, we remove the adjacency-related configuration segments from both neighbors; for \OmitAclRule errors, we remove the same line from from every device that contains a copy of the selected ACL; and for \uwmadison, we introduce the same error on a primary router and its corresponding backup router.


\begin{figure}
    \centering
    \begin{subfigure}{0.58\columnwidth}
        \includegraphics[width=1\columnwidth]{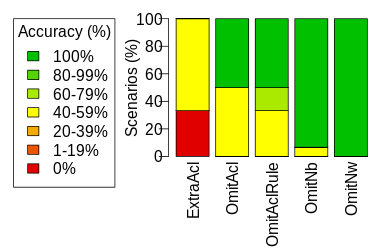}
        \caption{Recall}
        \label{fig:recall_scenario}
    \end{subfigure}
    \begin{subfigure}{0.38\columnwidth}
        \includegraphics[width=1\columnwidth]{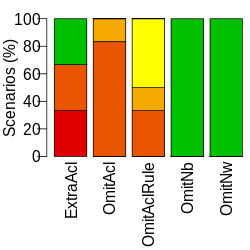}
        \caption{Precision}
        \label{fig:precision_scenario}
    \end{subfigure}
    \caption{Accuracy by error type}
    \label{fig:accuracy_scenario}
\end{figure}

\minisection{Impact of error type}
\figref{fig:recall_scenario} shows \name's recall for each type of error. \name locates all \OmitNetwork errors in all scenarios and all \OmitNeighbor errors in all but two scenarios. In contrast, for \OmitAclRule and \OmitAcl errors, \name locates all of the errors for \northwestern and about half of the errors for \uwmadison. As discussed above, \uwmadison replicates ACLs across devices, so we replicate ACL errors across devices. However, \name only identifies the errors on some devices. Nonetheless, this is sufficient to draw network engineer's attention to the problem. (We elaborate on differences in accuracy between networks below.) \name's recall is the worst for \AddAcl errors, because \name primarily enumerates MCSes containing interface state (up), routing adjacency, and layer-3 adjacency configuration segments. While disabling an interface or adjacency would satisfy the violated forwarding requirements, such changes would likely cause other forwarding requirements to be violated and not match engineers' management practices elsewhere in the network. Developing more sophisticated ranking heuristics that eschew these alternatives is an important area of future work.

\figref{fig:precision_scenario} shows \name's precision for each type of error. For all \OmitNetwork and \OmitNeighbor scenarios, \name only flags the configuration segments we omitted. In contrast, for ACL-related errors, \name's precision is generally less than 60\%. The cause of this low precision is the same as we discussed above for \AddAcl errors. This indicates that \name is better suited for routing-related errors; improving \name's accuracy for ACL errors and/or expanding the range of ACL errors other systems~\cite{jinjing, selfstarter} can detect is important future work.

\minisection{Impact of network design}
Next, we examine \name's accuracy for different networks. \name achieves perfect recall and precision for all synthetic configurations, except for one \OmitNeighbor scenario in one WAN. \aaron{Why?} \name's recall is also perfect for all \northwestern scenarios, but \name achieves perfect recall for only one-third of the \uwmadison scenarios and 50\% recall for most of the other \uwmadison scenarios. As we discussed above, \uwmadison replicates ACLs (and we replicate ACL errors) across devices, but \name only locates the errors on about half of the devices. If we increase the time limit on MCS enumeration (to 1 hour), then \name is able to explore more of the lattice (\secref{sec:mcses:compute:all}) and achieves higher recall.  \aaron{CHECK THIS!}

\name achieves 100\% precision for about half of the \uwmadison scenarios and about one-third of the \northwestern scenarios. In the remaining scenarios, which all contain ACL errors, the average precision is 25\% and 19\%, respectively. In summary, \name's accuracy is partially impacted by network design, but the type of error \name must locate has a greater impact.

\newcommand{\SmallestMCSes}{{\em {\name}Smallest}\xspace}
\newcommand{\ThreeSmallestMCSes}{{\em {\name}Three}\xspace}
\newcommand{\AllMCSes}{{\em All}\xspace}
\newcommand{\IntersectAcrossRequirements}{{\em Intersect}\xspace}

\begin{figure}
    \centering
    \begin{subfigure}{0.48\columnwidth}
        \includegraphics[width=1\columnwidth]{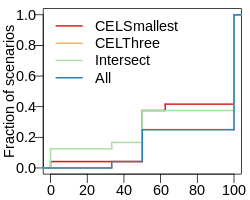}
        \caption{Recall}
        \label{fig:recall_rank}
    \end{subfigure}
    \begin{subfigure}{0.48\columnwidth}
        \includegraphics[width=1\columnwidth]{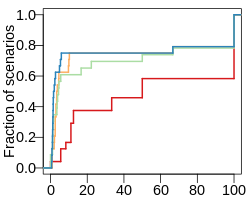}
        \caption{Precision}
        \label{fig:precision_rank}
    \end{subfigure}
    \caption{Impact of ranking}
    \label{fig:accuracy_rank}
\end{figure}

\minisection{Impact of ranking}
Next, we examine how \name's approach of aggregating MCSes from all violated forwarding requirements and ranking MCSes from smallest to largest (\secref{sec:mcses:rank}) impacts \name's accuracy. We compare four different approaches: \name's approach in which we only consider the smallest (i.e., highest ranked) MCSes (\SmallestMCSes), \name's approach in which consider the three smallest sizes of MCSes (\ThreeSmallestMCSes), only considering MCSes that are common across all violated forwarding requirements (\IntersectAcrossRequirements), and considering all MCSes \name enumerates within 10 minutes (\AllMCSes). We limit our analysis to real university configurations, because all scenarios with synthetic WAN configurations have only one MCS.

We observe (\figref{fig:recall_rank}) that \AllMCSes has perfect recall for 75\% of the scenarios, whereas \SmallestMCSes and \ThreeSmallestMCSes have perfect recall for 60\% of the scenarios. Furthermore, all approaches except \IntersectAcrossRequirements have at least 50\% recall for 96\% of the scenarios. However, when we look at precision (\figref{fig:precision_rank}), \SmallestMCSes has the best overall precision, with at least 33\% precision for 55\% of the scenarios and perfect precision for 40\% of the scenarios. In contrast, \AllMCSes has less than 6\% precision for 75\% of the scenarios and perfect precision for only 20\% of the scenarios; the precision for \ThreeSmallestMCSes is similar to \AllMCSes. In summary, \SmallestMCSes significantly improves precision without a substantial decrease in recall.

%% file: tables/synthetic_errors.tex
\begin{table}
\small
\centering
\begin{tabular}{c p{0.8\columnwidth}}
Type & Change $\rightarrow$ Impact \\
\hline
\OmitNetwork & Remove BGP or OSPF {\tt network} $\rightarrow$ \newline router does not advertise a prefix \\
\OmitNeighbor & Remove BGP {\tt neighbor} or OSPF {\tt no passive} $\rightarrow$ \newline router does not advertise routes to a neighbor \\
\OmitAcl & Remove {\tt ip access-group} from interface $\rightarrow$ \newline all packets are allowed \\
\OmitAclRule & Remove line from {\tt access-list} $\rightarrow$ \newline additional packets are blocked or allowed \\
\AddAcl & Add {\tt ip access-group} to interface $\rightarrow$ \newline some packets are blocked \\
\end{tabular}
\caption{Synthetic errors injected into configurations}
\label{tbl:synthetic_errors}
\end{table}

%% file: evaluation_efficiency.tex
\subsection{Efficiency}
\label{sec:evaluation:efficiency}

Lastly, we evaluate \name's efficiency. We measure the time \name takes to complete each of its major tasks: constructing the system of constraints (\secref{sec:encoding}), enumerating failure scenarios (\secref{sec:mcses:failures}), and enumerating MCSes (\secref{sec:mcses:compute}). We use the same configurations as \secref{sec:evaluation:accuracy}.
\name can be applied to multiple forwarding requirements in parallel, so the time required for a single forwarding requirement is the limiting factor. 

\begin{figure}
\centering
    \begin{subfigure}{0.48\columnwidth}
        \includegraphics[width=1\columnwidth]{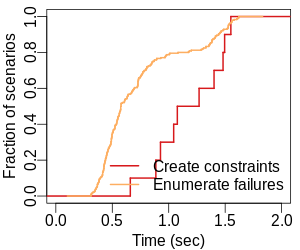}
        \caption{Constraints \& Failures}
        \label{fig:time_encoding_failures}
    \end{subfigure}
    \begin{subfigure}{0.48\columnwidth}
        \includegraphics[width=1\columnwidth]{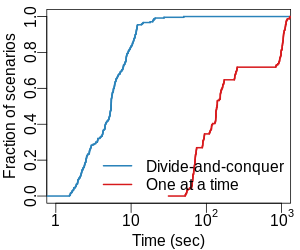}
        \caption{Generate a single MCS}
        \label{fig:time_mcses_single}
    \end{subfigure}
    \caption{Time to complete major tasks}
    \label{fig:time}
\end{figure}

We observe (\figref{fig:time_encoding_failures}) that \name can construct the system of constraints for the university and WAN networks in $\leq$1.5s. The time is strongly correlated (coeff = 0.70) with the number of routers in a network. 
Enumerating the failure scenarios under which a forwarding requirement is violated is also fast ($\leq$1.6s) for the networks and requirements we study.

Enumerating MCSes is the most time consuming task. For half of the scenarios, it takes $\leq$12s to enumerate all MCSes, while for 40\% of the scenarios, we reach the 10 minute time limit we set, and \name does not enumerate all MCSes. However, as we showed in \secref{sec:evaluation:accuracy}, \name achieves reasonable accuracy even with partial enumeration of MCSes.

As shown in \figref{fig:time_mcses_single}, our divide-and-conquer approach for computing MSSes (\secref{sec:mcses:compute:single}) decreases the time to compute a single MCS by approximately two orders of magnitude.


%% file: related.tex
\section{Related Work}

We discussed existing verification and synthesis tools in \secref{sec:techniques:research}.

Fault localization has been extensively studied by the software engineering research community~\cite{faultloc_survey}. \name's use of correction sets to identify faulty configuration statements is most closely related to Bug-Assist's use of a MaxSMT formulation to compute the maximal set of program statements that may cause a specific test case failure~\cite{causeclue, bugassist}. 
\name's aggregation of MCSes across failure scenarios and forwarding requirements (\secref{sec:mcses}) is inspired by spectrum-based fault localization techniques~\cite{tarantula}.


%% file: conclusion.tex
\section{Conclusion}
\label{sec:conclusion}

We have presented \name: a system that accurately localizes configuration errors by computing minimal correction sets (MCSes) for an SMT-based network model using domain-specific MCS enumeration algorithms. We showed that \name is able to pinpoint errors in real university networks and synthesized WAN configurations and identify all routing-related and half of all ACL-related synthetic errors we introduce.

%% file: encoding_appendix.tex
\section{Existing Network Models}
\label{sec:encoding:appendix}


As we briefly discussed in \secref{sec:encoding:background}, Minesweeper and NetComplete encode a network's forwarding behavior as a system of logical constraints over symbolic route advertisements. Symbolic advertisements are created for each routing adjacency---i.e., pair of neighboring routers that are configured to exchange routes using a particular routing protocol (e.g., OSPF). Each symbolic advertisement is composed of multiple symbolic variables that mirror the fields in actual route advertisements: e.g., prefix, path cost/length, and local preference. Constraints on these symbolic advertisements express: ($i$) the route advertisement import/export behavior of each router, ($ii$) the route selection procedures within each router, and ($iii$) the consequent packet forwarding behavior. This appendix explains each of these constraints in detail using the example network in \figref{fig:techniques_example}.


\minisection{Route import/export}
The encoding includes import and export constraints for each direction of a routing adjacency.
These constraints encode: ($i$) the application of route policies which filter and/or modify route advertisements, ($ii$) the forwarding of route advertisements, and ($iii$) the origination of route advertisements. For example, \figsref{fig:encoding_core1_import}{fig:encoding_core1_export} encode which OSPF advertisements \coreOneRouter imports from and exports to, respectively, \coreTwoRouter. A symbolic advertisement representing an imported route (e.g.,
$\incoming{OSPF}{\coreOneRouter}{\coreTwoRouter}$) 
is valid iff an advertisement is exported by the adjacent router (line 1 in \figref{fig:encoding_core1_import}) and the advertisement is accepted by the inbound route filter (true by default). A symbolic advertisement representing an exported route 
(e.g., $\outgoing{OSPF}{\coreOneRouter}{\coreTwoRouter}$) is valid iff the forwarding requirement's destination prefix falls within an originated prefix (line 1 in \figref{fig:encoding_core1_export}) or the best imported route (if any) is accepted by the outbound route filter (true by default) (line 4 in \figref{fig:encoding_core1_export}).


\input{algorithms/encoding_existing.tex}



\minisection{Route selection}
Protocol-specific route selection algorithms are encoded using constraints that compare the imported advertisements (e.g., $\incoming{OSPF}{\coreOneRouter}{\coreTwoRouter}$ and $\incoming{OSPF}{\coreOneRouter}{\coreThreeRouter}$) according to local preference, path length/cost, etc.  Cross-protocol route selection algorithms are encoded using constraints that compare the best advertisements from each protocol (e.g., \best{OSPF}{\coreOneRouter} and \best{BGP}{\coreOneRouter}) according to administrative distance. The highest-ranked symbolic advertisement (e.g., $\best{Overall}{\coreOneRouter}$) represents an entry in the router's global RIB.

\minisection{Packet forwarding}
Minesweeper also models the contents of a router's FIB. Constraints on symbolic forwarding variables encode the application of: ($i$) routes from the global RIB, and ($ii$) packet filters defined in configurations. For example, \figref{fig:encoding_core2_forward} encodes whether $\coreTwoRouter$ forwards to $\coreThreeRouter$. 

\minisection{Forwarding requirements}
A network's forwarding requirements are expressed in terms of constraints on symbolic route advertisements and/or symbolic forwarding entries in a manner similar to lines 1--3 in \figref{fig:challenges_encoding0}. 
\name considers one forwarding requirement at a time, but could easily be extended to handle multiple forwarding requirements at a time, as done by NetComplete.

Note that we have not discussed Minesweeper's encoding of link/node failure states (in route import/export constraints) or NetComplete's encoding of primary/backup paths (in forwarding requirements). We discuss \name's handling of failures in \secref{sec:mcses:failures}.

%% file: algorithms/encoding_existing.tex
\begin{figure}[t]
    \begin{subfigure}{1\columnwidth}
    \begin{algorithm}[H]
    \small
    \If(\Comment*[h]{Accept all exported}){$\outgoing{OSPF}{\coreTwoRouter}{\coreOneRouter}.\valid$}{
        $\incoming{OSPF}{\coreOneRouter}{\coreTwoRouter}.\valid = \true$\;
        $\incoming{OSPF}{\coreOneRouter}{\coreTwoRouter}.\prefix = \outgoing{OSPF}{\coreTwoRouter}{\coreOneRouter}.\prefix$\;
        $\incoming{OSPF}{\coreOneRouter}{\coreTwoRouter}.\cost = \outgoing{OSPF}{\coreTwoRouter}{\coreOneRouter}.\cost$\;
        }
    \lElse{$\incoming{OSPF}{\coreOneRouter}{\coreTwoRouter}.\valid = \false$ \Comment*[h]{None imported}}
    \end{algorithm}
    \caption{\coreOneRouter import from \coreTwoRouter}
    \label{fig:encoding_core1_import}
    \end{subfigure}
    
    \begin{subfigure}{1\columnwidth}
    \begin{algorithm}[H]
    \small
    \If(\Comment*[h]{Originate \deptOneSubnet}){$\destination \in 1.0.1.0/24$}{
        $\outgoing{OSPF}{\coreOneRouter}{\coreTwoRouter}.\prefix = 1.0.1.0/24$\;
        $\outgoing{OSPF}{\coreOneRouter}{\coreTwoRouter}.\cost = 1$\;
    }
    \ElseIf(\Comment*[h]{Forward shortest}){$\best{OSPF}{\coreOneRouter}.\valid$}{
        $\outgoing{OSPF}{\coreOneRouter}{\coreTwoRouter}.\valid = \true$\;
        $\outgoing{OSPF}{\coreOneRouter}{\coreTwoRouter}.\prefix = \best{OSPF}{\coreOneRouter}.\prefix$\;
        $\outgoing{OSPF}{\coreOneRouter}{\coreTwoRouter}.\cost = \best{OSPF}{\coreOneRouter}.\cost + 1$\;
    }
    \lElse{$\outgoing{OSPF}{\coreOneRouter}{\coreTwoRouter}.\valid = \false$ \Comment*[h]{None exported}}
    \end{algorithm}
    \caption{\coreOneRouter export to \coreTwoRouter}
    \label{fig:encoding_core1_export}
    \end{subfigure}
    
    \begin{subfigure}{1\columnwidth}
    \begin{algorithm}[H]
    \small
    $\forward{}{\coreTwoRouter}{\coreThreeRouter} \Leftrightarrow \best{Overall}{\coreTwoRouter} = \incoming{OSPF}{\coreTwoRouter}{\coreThreeRouter}$\;
    \Indp
    $\land\ (\source \in 1.0.2.0/24 \lor \source \in 1.0.3.0/24)$ \Comment*[h]{deptFilter}
    \end{algorithm}
    \caption{\coreTwoRouter forward to \coreThreeRouter}
    \label{fig:encoding_core2_forward}
    \end{subfigure}
    
    \caption{Constraints encoding part of the behavior of the example network in \figref{fig:techniques_example}}
    \label{fig:encoding_existing}
\end{figure}

%% file: main.bbl
\begin{thebibliography}{10}

\bibitem{ciscoiosxecommands}
Command reference, cisco ios xe amsterdam 17.2.x (catalyst 9600 switches).
\newblock
  \url{https://www.cisco.com/c/en/us/td/docs/switches/lan/catalyst9600/software/release/17-2/command_reference/b_172_9600_cr.html}.

\bibitem{tiramisu}
A.~Abhashkumar, A.~Gember-Jacobson, and A.~Akella.
\newblock Tiramisu: Fast multilayer network verification.
\newblock In {\em Symposium on Networked Systems Design and Implementation
  ({NSDI})}, 2020.

\bibitem{minesweeper}
R.~Beckett, A.~Gupta, R.~Mahajan, and D.~Walker.
\newblock A general approach to network configuration verification.
\newblock In {\em SIGCOMM}, 2017.

\bibitem{shapeshifter}
R.~Beckett, A.~Gupta, R.~Mahajan, and D.~Walker.
\newblock Abstract interpretation of distributed network control planes.
\newblock {\em Proceedings of the ACM on Programming Languages}, 4(POPL):1--27,
  2019.

\bibitem{propane}
R.~Beckett, R.~Mahajan, T.~Millstein, J.~Padhye, and D.~Walker.
\newblock Don't mind the gap: Bridging network-wide objectives and device-level
  configurations.
\newblock In {\em SIGCOMM}, 2016.

\bibitem{complexitymetrics}
T.~Benson, A.~Akella, and D.~Maltz.
\newblock Unraveling the complexity of network management.
\newblock In {\em Symposium on Networked Systems Design and Implementation
  ({NSDI})}, 2009.

\bibitem{adaptiveparsing}
D.~Caldwell, S.~Lee, and Y.~Mandelbaum.
\newblock Adaptive parsing of router configuration languages.
\newblock In {\em IEEE Internet Network Management Workshop ({INM})}, 2008.

\bibitem{diffprov}
A.~Chen, Y.~Wu, A.~Haeberlen, W.~Zhou, and B.~T. Loo.
\newblock The good, the bad, and the differences: Better network diagnostics
  with differential provenance.
\newblock In {\em SIGCOMM}, 2016.

\bibitem{dekleer}
J.~de~Kleer and B.~C. Williams.
\newblock Diagnosing multiple faults.
\newblock {\em Artif. Intell.}, 32(1):97--130, 1987.

\bibitem{netcomplete}
A.~El-Hassany, P.~Tsankov, L.~Vanbever, and M.~Vechev.
\newblock Netcomplete: Practical network-wide configuration synthesis with
  autocompletion.
\newblock In {\em NSDI}, 2018.

\bibitem{synet}
A.~El{-}Hassany, P.~Tsankov, L.~Vanbever, and M.~T. Vechev.
\newblock Network-wide configuration synthesis.
\newblock In {\em Computer Aided Verification ({CAV})}, 2017.

\bibitem{era}
S.~K. Fayaz, T.~Sharma, A.~Fogel, R.~Mahajan, T.~D. Millstein, V.~Sekar, and
  G.~Varghese.
\newblock Efficient network reachability analysis using a succinct control
  plane representation.
\newblock In {\em Symposium on Operating Systems Design and Implementation
  ({OSDI})}, 2016.

\bibitem{rcc}
N.~Feamster and H.~Balakrishnan.
\newblock Detecting {BGP} configuration faults with static analysis.
\newblock In {\em Symposium on Networked Systems Design and Implementation
  ({NSDI})}, 2005.

\bibitem{batfish}
A.~Fogel, S.~Fung, L.~Pedrosa, M.~Walraed-Sullivan, R.~Govindan, R.~Mahajan,
  and T.~Millstein.
\newblock A general approach to network configuration analysis.
\newblock In {\em Symposium on Networked Systems Design and Implementation
  ({NSDI})}, 2015.

\bibitem{cpr}
A.~Gember{-}Jacobson, A.~Akella, R.~Mahajan, and H.~H. Liu.
\newblock Automatically repairing network control planes using an abstract
  representation.
\newblock In {\em Symposium on Operating Systems Principles ({SOSP})}, 2017.

\bibitem{arc}
A.~Gember-Jacobson, R.~Viswanathan, A.~Akella, and R.~Mahajan.
\newblock Fast control plane analysis using an abstract representation.
\newblock In {\em SIGCOMM}, 2016.

\bibitem{mpa}
A.~Gember-Jacobson, W.~Wu, X.~Li, A.~Akella, and R.~Mahajan.
\newblock Management plane analytics.
\newblock In {\em Internet Measurement Conference ({IMC})}, 2015.

\bibitem{deltanet}
A.~Horn, A.~Kheradmand, and M.~R. Prasad.
\newblock Delta-net: Real-time network verification using atoms.
\newblock In {\em Symposium on Networked Systems Design and Implementation
  ({NSDI})}, 2017.

\bibitem{rcdc}
K.~Jayaraman, N.~Bj{\o}rner, J.~Padhye, A.~Agrawal, A.~Bhargava, P.~C.
  Bissonnette, S.~Foster, A.~Helwer, M.~Kasten, I.~Lee, A.~Namdhari, H.~Niaz,
  A.~Parkhi, H.~Pinnamraju, A.~Power, N.~M. Raje, and P.~Sharma.
\newblock Validating datacenters at scale.
\newblock In J.~Wu and W.~Hall, editors, {\em {SIGCOMM}}, 2019.

\bibitem{bugassist}
M.~Jose and R.~Majumdar.
\newblock Bug-assist: Assisting fault localization in {ANSI-C} programs.
\newblock In {\em Computer Aided Verification ({CAV})}, 2011.

\bibitem{causeclue}
M.~Jose and R.~Majumdar.
\newblock Cause clue clauses: error localization using maximum satisfiability.
\newblock In {\em {ACM} {SIGPLAN} Conference on Programming Language Design and
  Implementation ({PLDI})}, 2011.

\bibitem{selfstarter}
S.~K.~R. K., A.~Tang, R.~Beckett, K.~Jayaraman, T.~D. Millstein, Y.~Tamir, and
  G.~Varghese.
\newblock Finding network misconfigurations by automatic template inference.
\newblock In R.~Bhagwan and G.~Porter, editors, {\em 17th {USENIX} Symposium on
  Networked Systems Design and Implementation ({NSDI})}, 2020.

\bibitem{netplumber}
P.~Kazemian, M.~Chang, H.~Zeng, G.~Varghese, N.~McKeown, and S.~Whyte.
\newblock Real time network policy checking using header space analysis.
\newblock In {\em Symposium on Networked Systems Design and Implementation
  ({NSDI})}, 2013.

\bibitem{hsa}
P.~Kazemian, G.~Varghese, and N.~McKeown.
\newblock Header space analysis: Static checking for networks.
\newblock In {\em Symposium on Networked Systems Design and Implementation
  ({NSDI})}, 2012.

\bibitem{veriflow}
A.~Khurshid, X.~Zou, W.~Zhou, M.~Caesar, and P.~B. Godfrey.
\newblock {VeriFlow}: Verifying network-wide invariants in real time.
\newblock In {\em Symposium on Networked Systems Design and Implementation
  ({NSDI})}, 2013.

\bibitem{taleoftwocampuses}
H.~Kim, T.~Benson, A.~Akella, and N.~Feamster.
\newblock The evolution of network configuration: A tale of two campuses.
\newblock In {\em Internet Measurement Conference ({IMC})}, 2011.

\bibitem{topologyzoo}
S.~Knight, H.~X. Nguyen, N.~Falkner, R.~A. Bowden, and M.~Roughan.
\newblock The internet topology zoo.
\newblock {\em {IEEE} Journal on Selected Areas in Communications},
  29(9):1765--1775, 2011.

\bibitem{minerals}
F.~Le, S.~Lee, T.~Wong, H.~S. Kim, and D.~Newcomb.
\newblock Detecting network-wide and router-specific misconfigurations through
  data mining.
\newblock {\em {IEEE/ACM} Trans. Netw.}, 17(1):66--79, 2009.

\bibitem{marco}
M.~H. Liffiton, A.~Previti, A.~Malik, and J.~Marques{-}Silva.
\newblock Fast, flexible {MUS} enumeration.
\newblock {\em Constraints}, 21(2):223--250, 2016.

\bibitem{netcraft}
H.~H. Liu, X.~Wu, W.~Zhou, W.~Chen, T.~Wang, H.~Xu, L.~Zhou, Q.~Ma, and
  M.~Zhang.
\newblock Automatic life cycle management of network configurations.
\newblock In {\em SIGCOMM Workshop on Self-Driving Networks}, 2018.

\bibitem{bgpmisconfig}
R.~Mahajan, D.~Wetherall, and T.~E. Anderson.
\newblock Understanding {BGP} misconfiguration.
\newblock In {\em SIGCOMM}, 2002.

\bibitem{anteater}
H.~Mai, A.~Khurshid, R.~Agarwal, M.~Caesar, P.~B. Godfrey, and S.~T. King.
\newblock Debugging the data plane with {Anteater}.
\newblock In {\em SIGCOMM}, 2011.

\bibitem{flint}
N.~Narodytska, N.~Bj{\o}rner, M.~V. Marinescu, and M.~Sagiv.
\newblock Core-guided minimal correction set and core enumeration.
\newblock In {\em International Joint Conference on Artificial Intelligence
  ({IJCAI})}, 2018.

\bibitem{plankton}
S.~Prabhu, K.-Y. Chou, A.~Kheradmand, B.~Godfrey, and M.~Caesar.
\newblock Plankton: Scalable network configuration verification through model
  checking.
\newblock In {\em Symposium on Networked Systems Design and Implementation
  ({NSDI})}, 2019.

\bibitem{reiter}
R.~Reiter.
\newblock A theory of diagnosis from first principles.
\newblock {\em Artif. Intell.}, 32(1):57--95, 1987.

\bibitem{tarantula}
J.~R. Ruthruff, E.~Creswick, M.~M. Burnett, C.~R. Cook, S.~Prabhakararao, M.~F.
  II, and M.~Main.
\newblock End-user software visualizations for fault localization.
\newblock In {\em {ACM} 2003 Symposium on Software Visualization}, 2003.

\bibitem{zeppelin}
K.~Subramanian, L.~D'Antoni, and A.~Akella.
\newblock Synthesis of fault-tolerant distributed router configurations.
\newblock {\em {POMACS}}, 2(1):22:1--22:26, 2018.

\bibitem{jinjing}
B.~Tian, X.~Zhang, E.~Zhai, H.~H. Liu, Q.~Ye, C.~Wang, X.~Wu, Z.~Ji, Y.~Sang,
  M.~Zhang, et~al.
\newblock Safely and automatically updating in-network acl configurations with
  intent language.
\newblock In {\em {SIGCOMM}}, 2019.

\bibitem{codeomission}
X.~Wang.
\newblock {\em Automatic Localization of Code Omission Faults}.
\newblock PhD thesis, 2010.

\bibitem{bagpipe}
K.~Weitz, D.~Woos, E.~Torlak, M.~D. Ernst, A.~Krishnamurthy, and Z.~Tatlock.
\newblock Scalable verification of border gateway protocol configurations with
  an {SMT} solver.
\newblock In {\em {ACM} {SIGPLAN} International Conference on Object-Oriented
  Programming, Systems, Languages, and Applications ({OOPSLA})}, 2016.

\bibitem{faultloc_survey}
W.~E. Wong, R.~Gao, Y.~Li, R.~Abreu, and F.~Wotawa.
\newblock A survey on software fault localization.
\newblock {\em {IEEE} Trans. Software Eng.}, 42(8):707--740, 2016.

\bibitem{ynot}
Y.~Wu, M.~Zhao, A.~Haeberlen, W.~Zhou, and B.~T. Loo.
\newblock Diagnosing missing events in distributed systems with negative
  provenance.
\newblock In {\em SIGCOMM}, 2014.

\bibitem{z3}
The z3 theorem prover.
\newblock \url{https://github.com/Z3Prover/z3}.

\bibitem{atpg}
H.~Zeng, P.~Kazemian, G.~Varghese, and N.~McKeown.
\newblock Automatic test packet generation.
\newblock In {\em Conference on emerging Networking Experiments and
  Technologies ({CoNEXT})}, 2012.

\bibitem{apkeep}
P.~Zhang, X.~Liu, H.~Yang, N.~Kang, Z.~Gu, and H.~Li.
\newblock {APKeep}: Realtime verification for real networks.
\newblock In {\em NSDI}, 2020.

\end{thebibliography}
